\documentclass[prb,showpacs,showkeys]{revtex4}%
\usepackage{amssymb}
\usepackage{amsmath}
\usepackage{graphicx}
\usepackage{dcolumn}
\usepackage{bm}
\usepackage{amsfonts}
\usepackage{subfig}
\usepackage{float}%
\begin{document}
\title{Soliton states in mesoscopic two-band-superconducting cylinders}
\author{S. V. Kuplevakhsky}
\author{A. N. Omelyanchouk }
\author{Y. S. Yerin}
\email{yerin@ilt.kharkov.ua}
\affiliation{B. I. Verkin Institute for Low Temperature Physics and Engineering, }
\affiliation{National Academy of Sciences of Ukraine, }
\affiliation{47 Lenin Ave., 61103 Kharkiv, UKRAINE}
\date{\today }

\begin{abstract}
In the framework of the Ginzburg-Landau approach, we present a self-consistent
theory of specific soliton states in mesoscopic (thin-walled)
two-band-superconducting cylinders in external parallel magnetic fields. Such
states arise in the presence of "Josephson-type" interband coupling, when
phase winding numbers are different for each component of the superconducting
order parameter. We evaluate the Gibbs free energy of the sysyem up to
second-order terms in a certain dimensionless parameter $\varepsilon
\approx\frac{\mathcal{L}_{m}}{\mathcal{L}_{k}}\ll1$, where $\mathcal{L}_{m}$
and $\mathcal{L}_{k}~$are the magnetic and kinetic inductance, respectively.
We derive the complete set of exact soliton solutions. These solutions are
thoroughly analyzed from the viewpoint of both local and global
(thermodynamic) stability. In particular, we show that
rotational-symmetry-breaking caused by the formation of solitons gives rise to
a zero-frequency rotational mode. Although soliton states prove to be
thermodynamically metastable, the minimal energy gap between the lowest-lying
single-soliton states and thermodynamically stable zero-soliton states can be
much smaller than the magnetic Gibbs free energy of the latter states,
provided that intraband "penetration depths" differ substantially and
interband coupling is weak. The results of our investigation may apply to a
wide class of mesoscopic doubly-connected structures exhibiting two-band superconductivity.

\end{abstract}

\maketitle

\section{Introduction}

The subject of this paper is a self-consistent theory of specific soliton
states that were originally predicted in Ref. \cite{T02} and reportedly
observed experimentally.\cite{BKHM06} Without any doubt, these states can be
regarded as a hallmark of two-band superconductivity in mesoscopic
doubly-connected samples.

Indeed, owing to the emergence of additional degrees of freedom of the order
parameter, the nomenclature of topological objects in multiband
superconductors is much richer than that in conventional single-band
superconductors. In particular, Ginzburg-Landau equations describing two-band
superconductivity in bulk samples admit topologically stable solutions (with
one-dimensional singularities of the order parameter) that can be interpreted
as vortices carrying fractional magnetic flux.\cite{B02} In the absence of any
interband coupling, these vortices are accompanied by a circulating neutral
superfow associated with gradients of the interband phase difference. In the
presence of "Josephson-type" interband coupling, the neutral superflow
generates static solitons of the sine-Gordon type. In contrast to traditional
Abrikosov vortices in type-II superconductors, the energy per unit length of
these composite topological defects diverges at spatial infinity: hence they
are thermodynamically metastable and difficult to create in bulk samples.

However, solitons of the interband phase difference can exist by themselves in
doubly-connected mesoscopic samples, when the formation of any magnetic
vortices in the volume of the superconductor is prohibited
energetically.\cite{T02} Moreover, soliton states in this case can be induced
by an externally applied magnetic field, which makes them a convenient object
of investigation. Thus, experimental studies\cite{BKHM06} of the magnetic
response of mesoscopic two-band superconducting rings reveal certain
nontrivial features that, according to the authors of Ref. \cite{BKHM06}, can
be attributed to the creation of metastable soliton states. \qquad

Our research is largely motivated by the absence in current literature of any
quantitative theoretical analysis of this pronounced feature of two-band
superconductivity. (Unfortunately, the arguments of Ref. \cite{T02} and of the
recent publications\cite{TITWCST09} are mostly heuristic by nature.)
Mathematically, the approach of this paper is based on a Ginzburg-Landau-type
theory, which is a commonplace in theoretical studies of topological defects
in two-band superconductors: see the next section. This means, of course, that
we are restricted to the temperature range%
\[
\frac{T_{c}-T}{T_{c}}\ll1,
\]
where $T_{c}$ is the critical temperature of the superconducting transition.
\begin{figure}[ptb]
\includegraphics[height=0.5\textwidth]{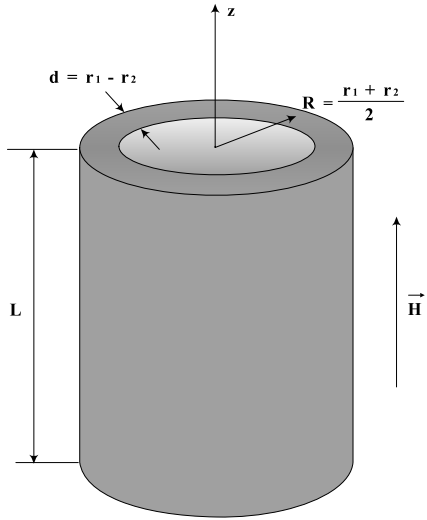} \caption{The geometry of the
problem (schematically). The parameters $L$, $R$ and $d$ obey conditions
(\ref{1.3})-(\ref{1.6}).}%
\label{fig:1}%
\end{figure}

As to the physical object, we consider a two-band superconductor in the form
of a straight, circular thin-walled cylinder, whose symmetry axis is the $z$
axis of cylindrical coordinates $\left(  r,\varphi,z\right)  $ (see Fig.
\ref{fig:1}). The constant external magnetic field $\mathbf{H}$ is applied
along the symmetry axis: $\mathbf{H}=\left(  0,0,H>0\right)  $. The length of
the generatrix of the wall of the cylinder satisfies the condition%
\begin{equation}
L\gg R\equiv\frac{r_{1}+r_{2}}{2}, \label{1.2}%
\end{equation}
which allows us to neglect end effects. The wall thickness $d\equiv
r_{2}-r_{1}$ and the average radius of the cylinder $R$ satisfy the following
conditions:%
\begin{equation}
d\ll\min\left\{  \xi_{1},\xi_{2}\right\}  , \label{1.3}%
\end{equation}%
\begin{equation}
R\gg\max\left\{  \xi_{1},\xi_{2}\right\}  , \label{1.4}%
\end{equation}%
\begin{equation}
R\gg\lambda, \label{1.5}%
\end{equation}%
\begin{equation}
\varepsilon\equiv\frac{dR}{2\lambda^{2}}\ll1, \label{1.6}%
\end{equation}
where $\xi_{1}$ and $\xi_{2}$ are the "coherence lengths" in bands 1 and 2,
respectively, and $\lambda$ is the weak-field penetration depth. Exact
definitions of $\xi_{1}$, $\xi_{2}$, and $\lambda$ will be given in the next
section; however, the role of conditions (\ref{1.3})- (\ref{1.6}) should be
explained right now. Thus, condition (\ref{1.3}) precludes the formation of
any magnetic vortices in the wall of the cylinder. Condition (\ref{1.4})
mainly simplifies mathematical consideration. In contrast, a combination of
conditions (\ref{1.5}) and (\ref{1.6}) is of primary importance: taken
together, these two conditions guarantee that self-induced magnetic fields are
small and can be treated perturbatively. (This fact justifies the definition
"mesoscopic cylinders" in the title of the paper.) Moreover, as will be shown,
the dimensionless quantity $\varepsilon$ serves as a natural expansion
parameter of the Gibbs free energy. In order to carry out a rigorous analysis
of thermodynamic stability of soliton states, we will have to evaluate the
Gibbs free energy \textit{exactly} up to small terms of order $\varepsilon
^{2}$, which implies the necessity of self-consistent evaluation of the vector
potential up to first-order terms in $\varepsilon$.

We conclude the formulation of the problem by specifying boundary conditions
for soliton states. Consider a two-component superconducting order parameter
$\Psi=\left(  \psi_{1},\psi_{2}\right)  $, where $\psi_{1}=\left\vert \psi
_{1}\right\vert e^{i\phi_{1}}$ and $\psi_{2}=\left\vert \psi_{2}\right\vert
e^{i\phi_{2}}$. The double-connectedness of the cylinder is accounted for by
the condition of single-valuedness of $\Psi$.\cite{M62} In particular, this
condition requires that%
\begin{equation}
\left.  \left\vert \psi_{1}\right\vert \right\vert _{\varphi=0}=\left.
\left\vert \psi_{1}\right\vert \right\vert _{\varphi=2\pi},\quad\left.
\left\vert \psi_{2}\right\vert \right\vert _{\varphi=0}=\left.  \left\vert
\psi_{2}\right\vert \right\vert _{\varphi=2\pi}. \label{1.7}%
\end{equation}
As to the phases $\phi_{1}$ and $\phi_{2}$, the requirement is as follows:%
\begin{equation}%
%TCIMACRO{\doint \nolimits_{\Gamma}}%
%BeginExpansion
{\displaystyle\oint\nolimits_{\Gamma}}
%EndExpansion
\nabla\phi_{1}\cdot d\mathbf{l}=2\pi n_{1},\quad%
%TCIMACRO{\doint \nolimits_{\Gamma}}%
%BeginExpansion
{\displaystyle\oint\nolimits_{\Gamma}}
%EndExpansion
\nabla\phi_{2}\cdot d\mathbf{l}=2\pi n_{2},\quad n_{1,2}=0,\pm1,\pm2,\ldots,
\label{1.8}%
\end{equation}
where $\Gamma$ is an arbitrary closed continuous contour that lies inside the
wall of the cylinder and encircles the opening. It should be emphasized that
there are \textit{no} a priori reasons for setting $n_{1}=n_{2}$.\cite{M79} As
in the case of fractional magnetic vortices in bulk two-band
superconductors,\cite{B02} nontrivial topological states arise when $n_{1}\neq
n_{2}$. In the presence of interband coupling, they are of the soliton type.

In Section II, we introduce the Gibbs free-energy functional of the system and
analyze its basic properties. In Section III, we derive a self-consistent
expression for the electromagnetic Gibbs free energy. Soliton solutions are
derived and thoroughly discussed in Section IV. Finally, in Section V, we
summarize the obtained results and make several concluding remarks. Appendices
A and B contain details of some mathematical calculations skipped over in the
main text. In Appendix C, we present several particular examples of soliton solutions.

\section{Gibbs free-energy functional}

We begin by defining the Gibbs free-energy functional of the system. In view
of complete homogeneity along the $z$ axis and with the normal-state Gibbs
free energy being subtracted, it takes the following form:%
\begin{gather}
G\left[  \Psi,\Psi^{\ast},\mathbf{A};\mathbf{H}\right]  =L%
%TCIMACRO{\dint \nolimits_{\Sigma_{S}}}%
%BeginExpansion
{\displaystyle\int\nolimits_{\Sigma_{S}}}
%EndExpansion
d^{2}\mathbf{r}\left[  \alpha_{1}\left\vert \psi_{1}\right\vert ^{2}%
+\alpha_{2}\left\vert \psi_{2}\right\vert ^{2}+\frac{\beta_{1}}{2}\left\vert
\psi_{1}\right\vert ^{4}+\frac{\beta_{2}}{2}\left\vert \psi_{2}\right\vert
^{4}\right. \nonumber\\
\left.  +\frac{1}{2m_{1}}\left\vert \left(  -i\hbar\nabla-\frac{2e}%
{c}\mathbf{A}\right)  \psi_{1}\right\vert ^{2}+\frac{1}{2m_{2}}\left\vert
\left(  -i\hbar\nabla-\frac{2e}{c}\mathbf{A}\right)  \psi_{2}\right\vert
^{2}-\gamma\left(  \psi_{1}\psi_{2}^{\ast}+\psi_{1}^{\ast}\psi_{2}\right)
\right] \nonumber\\
+\frac{L}{8\pi}%
%TCIMACRO{\dint \nolimits_{\Sigma_{S}+\Sigma_{O}}}%
%BeginExpansion
{\displaystyle\int\nolimits_{\Sigma_{S}+\Sigma_{O}}}
%EndExpansion
d^{2}\mathbf{r}\left(  \mathbf{h}-\mathbf{H}\right)  ^{2}. \label{2.1}%
\end{gather}
Here, the coefficients $\beta_{1}$ and $\beta_{2}$ are positive constants,
whereas $\alpha_{1}$ and $\alpha_{2}$ are temperature-dependent:%
\begin{equation}
\alpha_{1}=\alpha_{1}\left(  T\right)  \equiv a_{1}\left(  T-T_{1}\right)
,\quad\alpha_{1}=\alpha_{2}\left(  T\right)  \equiv a_{2}\left(
T-T_{2}\right)  ,\quad a_{1},a_{2},T_{1},T_{2}>0. \label{2.2}%
\end{equation}
Moreover, the latter coefficients enter the definitions of the coherence
lengths $\xi_{1}$ and $\xi_{2}$:%
\begin{equation}
\xi_{1}=\frac{\hbar}{\sqrt{2m_{1}\alpha_{1}}},\quad\xi_{2}=\frac{\hbar}%
{\sqrt{2m_{2}\alpha_{2}}}. \label{2.3}%
\end{equation}
The electron charge in (\ref{2.1}) is $e<0$; the total vector potential
$\mathbf{A}$ defines the local magnetic field $\mathbf{h}$:%
\begin{equation}
\mathbf{h}=\nabla\times\mathbf{A},\quad\mathbf{h}=\left(  0,0,h\right)  ,\quad
h=h\left(  r\right)  . \label{2.4}%
\end{equation}
The parameter of interband coupling, $\gamma$, may have either sign.
Two-dimensional integration in the plane $\left(  r,\varphi\right)  $ is
carried out over the cross-section of the superconductor ($\Sigma_{S}$) in the
square-bracketed terms, and over the cross-sections of the superconductor and
of the opening ($\Sigma_{S}+\Sigma_{O}$) in the last (magnetic) term.

A microscopic derivation of free-energy functionals of the type (\ref{2.1})
was given in Ref. \cite{Zh04}\ for the case of clean two-band superconductors
in the limit of small interband coupling. Free-energy functionals of this type
are employed in theoretical studies of different aspects of two-band
superconductivity, such as, e.g., topological defects,\cite{T02,B02,GV03}
current-carrying states,\cite{YO07} the Little-Parks effect,\cite{YKO08}
surface energy,\cite{GFKS10} etc. It should be additionally noted that, for
our specific geometry and $\gamma>0$, the functional (\ref{2.1}) also applies
to the description of a composite system consisting of two thin coaxial
cylindrical films of single-band superconductors, Josephson coupled via a
parallel insulating layer,\cite{KN97} which is exactly the experimental set-up
of Ref. \cite{BKHM06}.

To obtain the actual (observable) Gibbs free energy, one has to minimize
(\ref{2.1}) with respect to $\Psi$, $\Psi^{\ast}$ and $\mathbf{A}$ under
appropriate boundary conditions; however, substantial simplifications can be
made already at this stage. First, we notice that, by the symmetry of the
problem, the amplitudes $\left\vert \psi_{1}\right\vert $ and $\left\vert
\psi_{2}\right\vert $ do not depend on $\varphi$. Moreover, they cannot depend
on $r$, either. Indeed, by virtue of condition (\ref{1.3}), any radial
variations of $\left\vert \psi_{1}\right\vert $ and $\left\vert \psi
_{2}\right\vert $ would give rise to free-energy terms that are by the factors
$\frac{\xi_{1}^{2}}{d^{2}}\gg1$ and $\frac{\xi_{2}^{2}}{d^{2}}\gg1$ larger
than the first and the second terms in (\ref{2.1}), respectively, which is
energetically unfavorable.\cite{dG66} As a result, in equilibrium, the
magnitudes $\left\vert \psi_{1}\right\vert $ and $\left\vert \psi
_{2}\right\vert $ are functions of $T$\ and $H$ only.

Consider now the kinetic-energy terms [the first two terms in the second line
of (\ref{2.1})]. The ratio of these terms to the first and the second terms in
(\ref{2.1}), respectively, is at most of order $\frac{\xi_{1}^{2}}{R^{2}%
}\left(  \frac{\Phi_{H}}{\Phi_{0}}\right)  ^{2}$ and $\frac{\xi_{2}^{2}}%
{R^{2}}\left(  \frac{\Phi_{H}}{\Phi_{0}}\right)  ^{2}$, where $\Phi_{H}$ is
the external flux, and
\begin{equation}
\Phi_{0}=\frac{\pi\hbar c}{\left\vert e\right\vert } \label{2.5}%
\end{equation}
is the flux quantum. Owing to condition (\ref{1.4}), for sufficiently weak
external fields, $\frac{\xi_{1}^{2}}{R^{2}}\left(  \frac{\Phi_{H}}{\Phi_{0}%
}\right)  ^{2},\frac{\xi_{2}^{2}}{R^{2}}\left(  \frac{\Phi_{H}}{\Phi_{0}%
}\right)  ^{2}\ll1$. (Compare with the consideration of flux quantization in
singe-band-superconducting cylinders.\cite{G62}) In this field range, we can
set $\left\vert \psi_{1}\right\vert =\left\vert \psi_{1}\right\vert _{0}$ and
$\left\vert \psi_{2}\right\vert =\left\vert \psi_{2}\right\vert _{0}$, where
$\left\vert \psi_{1}\right\vert _{0}$ and $\left\vert \psi_{2}\right\vert
_{0}$ satisfy the equilibrium conditions for an unperturbed two-band
superconductor:
\begin{align}
\alpha_{1}\left\vert \psi_{1}\right\vert _{0}+\beta_{1}\left\vert \psi
_{1}\right\vert _{0}^{3}-\left\vert \gamma\right\vert \left\vert \psi
_{2}\right\vert _{0}  &  =0,\nonumber\\
\alpha_{2}\left\vert \psi_{2}\right\vert _{0}+\beta_{2}\left\vert \psi
_{2}\right\vert _{0}^{3}-\left\vert \gamma\right\vert \left\vert \psi
_{1}\right\vert _{0}  &  =0. \label{2.6}%
\end{align}
One can readily obtain a good approximate solution to (\ref{2.6}).\cite{A06}
However, it is of no interest in the context of the soliton problem. We only
note that the critical temperature, derived from (\ref{2.6}), is $T_{c}%
=\frac{1}{2}\left[  T_{1}+T_{2}+\sqrt{\left(  T_{1}-T_{2}\right)  ^{2}%
+\frac{\gamma^{2}}{a_{1}a_{2}}}\right]  $.

In light of these simplifications, it is reasonable to consider the weak-field
penetration depth\cite{A06}%
\begin{equation}
\lambda=\frac{c}{4\sqrt{\pi}\left\vert e\right\vert }\sqrt{\frac{m_{1}m_{2}%
}{m_{2}\left\vert \psi_{1}\right\vert _{0}^{2}+m_{1}\left\vert \psi
_{2}\right\vert _{0}^{2}}} \label{2.8}%
\end{equation}
and to define intraband "penetration depths"\cite{GFKS10}%
\begin{equation}
\lambda_{1}=\frac{c}{4\sqrt{\pi}\left\vert e\right\vert }\frac{\sqrt{m_{1}}%
}{\left\vert \psi_{1}\right\vert _{0}},\quad\lambda_{2}=\frac{c}{4\sqrt{\pi
}\left\vert e\right\vert }\frac{\sqrt{m_{2}}}{\left\vert \psi_{2}\right\vert
_{0}};\quad\lambda_{1}^{-2}+\lambda_{2}^{-2}=\lambda^{-2}. \label{2.9}%
\end{equation}
(For the above-mentioned composite, Josephson-coupled system, the quantities
$\lambda_{1}$ and $\lambda_{2}$ have direct physical meaning.) We also
introduce new, functionally independent phase variables $\not \phi $ and
$\chi$:\cite{YO07,YKO08}%
\begin{gather}
\phi=\varphi_{1}-\varphi_{2},\label{2.10}\\
\chi=c_{1}\varphi_{1}+c_{2}\varphi_{2};\quad c_{1}\equiv\left(  \lambda
\lambda_{1}^{-1}\right)  ^{2},\quad c_{2}\equiv\left(  \lambda\lambda_{2}%
^{-1}\right)  ^{2},\quad c_{1}+c_{2}=1. \label{2.11}%
\end{gather}

Using definitions (\ref{2.5}) and (\ref{2.8})-(\ref{2.11}), we obtain the
reduced Gibbs free-energy functional in the following form:%
\begin{equation}
G\left[  \phi,\chi\,,\mathbf{A};\mathbf{H}\right]  =F_{S0}+G_{em}\left[
\chi\,,\mathbf{A};\mathbf{H}\right]  +F_{sol}\left[  \phi\right]  .
\label{2.12}%
\end{equation}
Here, the first term is the free energy of the unperturbed superconducting
cylinder:\cite{YKO08}%
\begin{gather}
F_{S0}=V_{S}\left(  \alpha_{1}\left\vert \psi_{1}\right\vert _{0}^{2}%
+\alpha_{2}\left\vert \psi_{2}\right\vert _{0}^{2}+\frac{\beta_{1}}%
{2}\left\vert \psi_{1}\right\vert _{0}^{4}+\frac{\beta_{2}}{2}\left\vert
\psi_{2}\right\vert _{0}^{4}-2\left\vert \gamma\right\vert \left\vert \psi
_{1}\right\vert _{0}\left\vert \psi_{2}\right\vert _{0}\right)  ;
\label{2.13}\\
V_{S}\equiv2\pi RLd.\nonumber
\end{gather}
The second term is the electromagnetic Gibbs free-energy functional:%
\begin{equation}
G_{em}\left[  \chi\,,\mathbf{A};\mathbf{H}\right]  =\frac{\Phi_{0}^{2}}%
{32\pi^{3}}\frac{L}{\lambda^{2}}%
%TCIMACRO{\dint \nolimits_{\Sigma_{S}}}%
%BeginExpansion
{\displaystyle\int\nolimits_{\Sigma_{S}}}
%EndExpansion
d^{2}\mathbf{r}\left(  \nabla\chi-\frac{2e}{\hbar c}\mathbf{A}\right)
^{2}+\frac{L}{8\pi}%
%TCIMACRO{\dint \nolimits_{\Sigma_{S}+\Sigma_{O}}}%
%BeginExpansion
{\displaystyle\int\nolimits_{\Sigma_{S}+\Sigma_{O}}}
%EndExpansion
d^{2}\mathbf{r}\left(  \mathbf{h}-\mathbf{H}\right)  ^{2}, \label{2.14}%
\end{equation}
with the first term on the right-hand side of (\ref{2.14}) being the
kinetic-energy functional of the supercurrent. Finally, the last term in
(\ref{2.12}) is%
\begin{gather}
F_{sol}\left[  \phi\right]  =\frac{\Phi_{0}^{2}}{32\pi^{3}}\frac{L}%
{\lambda^{2}}c_{1}c_{2}%
%TCIMACRO{\dint \nolimits_{\Sigma_{S}}}%
%BeginExpansion
{\displaystyle\int\nolimits_{\Sigma_{S}}}
%EndExpansion
d^{2}\mathbf{r}\left[  \left(  \nabla\phi\right)  ^{2}+\frac{2}{l^{2}}\left(
1-\text{sgn~}\gamma\cos\phi\right)  \right]  ;\label{2.15}\\
l^{2}\equiv\frac{\Phi_{0}^{2}}{32\pi^{3}}\frac{1}{\lambda^{2}}\frac{c_{1}%
c_{2}}{\left\vert \gamma\right\vert \left\vert \psi_{1}\right\vert
_{0}\left\vert \psi_{2}\right\vert _{0}},\nonumber
\end{gather}
where sgn~$x$ is the sign function. The term (\ref{2.15}) should be
interpreted as the soliton self-energy functional. Indeed, when $n_{1}=n_{2}$
in (\ref{1.8}), we have\cite{YO07,YKO08} either $\phi=0$ mod $2\pi$ (for
$\gamma>0$) or $\phi=\pi$ mod $2\pi$ (for $\gamma<0$), and this term vanishes identically.

Our task now is to minimize (\ref{2.12}) with respect to $\phi$, $\chi$ and
$\mathbf{A}$. As the phase variable $\phi$ is not coupled to the vector
potential $\mathbf{A}$, this procedure can be performed in two separate steps.

\section{Electromagnetic Gibbs free energy}

The minimization of the electromagnetic functional (\ref{2.14}) reduces to
evaluation of the stationarity condition $\delta G_{em}=0$, or%
\begin{equation}
\frac{\delta G_{em}}{\delta\mathbf{A}}=0,\quad\frac{\delta G_{em}}{\delta\chi
}=0. \label{2.17}%
\end{equation}
Indeed, in view of quadratic nature of (\ref{2.14}), solutions to (\ref{2.17})
are automatically minimizers of this functional (i.e., $\delta^{2}G_{em}>0 $
at these solutions).

Variation with respect to $\mathbf{A}$ yields Amp\`{e}re's law%
\begin{gather}
\nabla\times\nabla\times\mathbf{A}=0,\quad r\in\left(  0,r_{1}\right)
;\label{2.18}\\
\nabla\times\nabla\times\mathbf{A}=\frac{4\pi}{c}\mathbf{j},\quad r\in\left(
r_{1},r_{2}\right)  , \label{2-19}%
\end{gather}
with%
\begin{equation}
\mathbf{j}=-\frac{c}{4\pi\lambda^{2}}\left(  \frac{\Phi_{0}}{2\pi}\nabla
\chi+\mathbf{A}\right)  \label{2.20}%
\end{equation}
being the supercurrent density [$\mathbf{j}=\left(  0,j,0\right)  $ by
symmetry], and the boundary condition%
\begin{equation}
\left.  \mathbf{h}\right\vert _{r=r_{2}}\equiv\left.  \nabla\times
\mathbf{A}\right\vert _{r=r_{2}}=\mathbf{H}. \label{2.21}%
\end{equation}
(This boundary condition should, of course, be complemented by the conditions
of continuity of $\mathbf{A}$ and $\mathbf{h}$ at $r=r_{2}$ and the condition
of regularity of $\mathbf{A}$ at the origin.) Variation with respect to $\chi
$, under the condition of single-valuedness of variations $\delta\chi$, just
yields the current-conservation law%
\begin{equation}
\nabla\mathbf{j}=0 \label{2.22}%
\end{equation}
and the single-valuedness condition%
\begin{equation}
\left.  j\right\vert _{\varphi=0}=\left.  j\right\vert _{\varphi=2\pi}.
\label{2.23}%
\end{equation}
[This boundary condition should be complemented by a condition on $\chi$
resulting from (\ref{1.8}).]

The problem of finding $\mathbf{A}$ and $\chi$ is still sub-definite, because
we have not so far fixed the gauge. As the $z$ component of the vector
potential drops out of the definition of $\mathbf{h}$ [see (\ref{2.4})], it is
equal to an arbitrary constant, and we set $A_{z}\equiv0$. The $r$ component
of the vector potential can be eliminated by the gauge transformation%
\[
\mathbf{A}\rightarrow\mathbf{A}-\nabla%
%TCIMACRO{\dint \nolimits_{0}^{r}}%
%BeginExpansion
{\displaystyle\int\nolimits_{0}^{r}}
%EndExpansion
A_{r}\left(  r^{\prime},\varphi\right)  dr^{\prime},\quad\chi\rightarrow
\chi+\frac{2\pi}{\Phi_{0}}%
%TCIMACRO{\dint \nolimits_{0}^{r}}%
%BeginExpansion
{\displaystyle\int\nolimits_{0}^{r}}
%EndExpansion
A_{r}\left(  r^{\prime},\varphi\right)  dr^{\prime}.
\]
In this particular gauge,%
\begin{gather}
\mathbf{A}=\left(  0,A,0\right)  ,\quad A=A\left(  r\right)  ;\label{2.25}\\
h=\frac{1}{r}\frac{d}{dr}\left(  rA\right)  , \label{2.26}%
\end{gather}
and $\chi$ does not depend on $r$ ($j_{r}\equiv0$). Using (\ref{1.8}),
(\ref{2.11}), (\ref{2.22}) and (\ref{2.23}), we arrive at a well-posed
boundary-value problem,%
\begin{gather*}
\frac{d^{2}\chi}{d\varphi^{2}}=0,\quad\varphi\in\left(  0,2\pi\right)  ;\\
\chi\left(  2\pi\right)  =\chi\left(  0\right)  +2\pi\left(  n_{1}c_{1}%
+n_{2}c_{2}\right)  ,\quad\frac{d\chi}{d\varphi}\left(  2\pi\right)
=\frac{d\chi}{d\varphi}\left(  0\right)  ,
\end{gather*}
whose solution is%
\begin{equation}
\chi\left(  \varphi\right)  =\left(  n_{1}c_{1}+n_{2}c_{2}\right)
\varphi+\varphi_{0}, \label{2.27}%
\end{equation}
with $\varphi_{0}$ being an arbitrary constant.

The boundary-value problem for the vector potential now takes the form%
\begin{gather}
\frac{d}{dr}\left[  \frac{1}{r}\frac{d}{dr}\left(  rA\right)  \right]
=0,\quad r\in\left(  0,r_{1}\right)  ;\nonumber\\
\frac{d}{dr}\left[  \frac{1}{r}\frac{d}{dr}\left(  rA\right)  \right]
=\frac{1}{\lambda^{2}}\left[  A+\frac{\Phi_{0}}{2\pi}q\left(  n_{1}%
,n_{2}\right)  \right]  ,\quad r\in\left(  r_{1},r_{2}\right)  ;\nonumber\\
q\left(  n_{1},n_{2}\right)  \equiv n_{1}c_{1}+n_{2}c_{2};\label{2.28}\\
\left.  \left\vert A\right\vert \right\vert _{r=0}<\infty,\,\left.
A\right\vert _{r=r_{1}-0}=\left.  A\right\vert _{r=r_{1}+0},\,\left.  \frac
{1}{r}\frac{d}{dr}\left(  rA\right)  \right\vert _{r=r_{1}-0}=\left.  \frac
{1}{r}\frac{d}{dr}\left(  rA\right)  \right\vert _{r=r_{1}+0},\nonumber\\
\left.  \frac{1}{r}\frac{d}{dr}\left(  rA\right)  \right\vert _{r=r_{2}%
}=H.\nonumber
\end{gather}
This boundary-value problem admits an exact solution: it is presented in
Appendix A. However, to obtain a second-order expansion of the electromagnetic
Gibbs free energy in terms of the small parameter $\varepsilon$ (see
Introduction), we need only first-order expansions of $A$ and $h$. They are as
follows:%
\begin{align}
A\left(  r\right)   &  =\frac{r}{2}H-\frac{r}{2}\left[  \frac{\Phi_{0}}{\pi
r_{1}^{2}}q\left(  n_{1},n_{2}\right)  +H\right]  \varepsilon,\quad
r\in\left[  0,r_{1}\right]  ;\nonumber\\
&  =\frac{r}{2}H-\frac{r_{1}}{2}\left[  \frac{\Phi_{0}}{\pi r_{1}^{2}}q\left(
n_{1},n_{2}\right)  +H\right]  \varepsilon,\quad r\in\left(  r_{1}%
,r_{2}\right]  ; \label{2.29}%
\end{align}%
\begin{align}
h\left(  r\right)   &  =H-\left[  \frac{\Phi_{0}}{\pi r_{1}^{2}}q\left(
n_{1},n_{2}\right)  +H\right]  \varepsilon,\quad r\in\left[  0,r_{1}\right]
;\nonumber\\
&  =H-\frac{r_{2}-r}{r_{2}-r_{1}}\left[  \frac{\Phi_{0}}{\pi r_{1}^{2}%
}q\left(  n_{1},n_{2}\right)  +H\right]  \varepsilon,\quad r\in\left(
r_{1},r_{2}\right]  . \label{2.30}%
\end{align}
[The fact that expressions (\ref{2.29}) and (\ref{2.30}) on the interval
$\left(  r_{1},r_{2}\right)  $ are \textit{not} related to each other by
equation (\ref{2.26}) should not cause any confusion: to ensure the
fulfillment of (\ref{2.26}), we would have to continue the expansion of
(\ref{a4}) up to small terms of order $\frac{d}{R}$ and $\varepsilon\frac
{d}{R}$.]

Bearing in mind that in integral physical quantities any difference between
$r_{1}$, $r_{2}$ and $R$ should be neglected [see (\ref{a8})], by use of
equations (\ref{2.30}) and (\ref{2-19}) we immediately obtain expressions for
the total flux $\Phi=%
%TCIMACRO{\dint \nolimits_{\Sigma_{O}}}%
%BeginExpansion
{\displaystyle\int\nolimits_{\Sigma_{O}}}
%EndExpansion
hd^{2}\mathbf{r}$ and the total supercurrent $J=L%
%TCIMACRO{\dint \nolimits_{r_{1}}^{r_{2}}}%
%BeginExpansion
{\displaystyle\int\nolimits_{r_{1}}^{r_{2}}}
%EndExpansion
jdr$:%
\begin{equation}
\Phi=\Phi_{H}+\Phi_{i},\quad\Phi_{H}=\pi R^{2}H,\quad\Phi_{i}=-\left[
\Phi_{0}q\left(  n_{1},n_{2}\right)  +\Phi_{H}\right]  \varepsilon;
\label{2.31}%
\end{equation}%
\begin{equation}
J=\frac{c}{\mathcal{L}_{m}}\Phi_{i},\quad\mathcal{L}_{m}=\frac{4\pi^{2}R^{2}%
}{L}. \label{2.32}%
\end{equation}
Here, $\Phi_{H}$ and $\Phi_{i}$ are the external and self-induced flux,
respectively; $\mathcal{L}_{m}$ is the magnetic inductance of the cylinder (or
self-inductance).\cite{LL68}

On substitution of relations (\ref{2.27}), (\ref{2.29}) and (\ref{2.30}) into
(\ref{2.14}), we get the electromagnetic Gibbs free energy of the state
parameterized by topological numbers $n_{1}$ and $n_{2}$:%
\begin{gather}
G_{em}\left(  n_{1},n_{2};H\right)  =\frac{\left[  \Phi_{0}q\left(
n_{1},n_{2}\right)  +\Phi_{H}\right]  ^{2}}{2\mathcal{L}_{m}}\varepsilon
\left(  1-2\varepsilon\right)  +\frac{\left[  \Phi_{0}q\left(  n_{1}%
,n_{2}\right)  +\Phi_{H}\right]  ^{2}}{2\mathcal{L}_{m}}\varepsilon
^{2}\nonumber\\
=\frac{\Phi_{0}^{2}\varepsilon}{2\mathcal{L}_{m}}\left[  q\left(  n_{1}%
,n_{2}\right)  +f_{H}\right]  ^{2}\left(  1-\varepsilon\right)  ,\quad
f_{H}\equiv\frac{\Phi_{H}}{\Phi_{0}}. \label{2.33}%
\end{gather}
(Note that the first term in the first line of this equation is the kinetic
energy of the supercurrent, whereas the second term in the same line is the
magnetic Gibbs free energy.)

To facilitate an analysis of soliton states, we should transform (\ref{2.33})
to a more convenient form. First, instead of parameterization by $n_{1}$ and
$n_{2}$, we introduce parameterization by $n_{1}$ and the soliton number
$n=n_{1}-n_{2}$. From now on, we assume (without any loss of generality) that
$0<c_{2}\leq c_{1}<1$. Relation (\ref{2.33}) is rewritten as follows:%
\begin{equation}
G_{em}\left(  n_{1},n;f_{H}\right)  =\frac{\Phi_{0}^{2}\varepsilon
}{2\mathcal{L}_{m}}\left(  f_{H}+n_{1}-nc_{2}\right)  ^{2}\left(
1-\varepsilon\right)  . \label{2.34}%
\end{equation}

From a thermodynamic point of view, of interest is the minimum of (\ref{2.34})
for given $\left\vert n\right\vert =0,1,2,\ldots$.Therefore, relation
(\ref{2.34}) should be minimized with respect to $n_{1}$ and sng$~n$. With
this in mind, we introduce two discontinuous functions, a step function
$m\left(  x\right)  $ and a periodic function $\theta\left(  x\right)  $, via
the definitions%
\begin{equation}
m\left(  x\right)  =\left\{
\begin{array}
[c]{c}%
\left[  x\right]  ,\quad0\leq\left\{  x\right\}  \leq\frac{1}{2};\\
\left[  x\right]  +1,\quad\frac{1}{2}<\left\{  x\right\}  <1,
\end{array}
\right.  \label{2.35}%
\end{equation}
and%
\begin{equation}
\theta\left(  x\right)  =\left\{
\begin{array}
[c]{c}%
\left\{  x\right\}  ,\quad0\leq\left\{  x\right\}  \leq\frac{1}{2};\\
-1+\left\{  x\right\}  ,\quad\frac{1}{2}<\left\{  x\right\}  <1,
\end{array}
\right.  \label{2.36}%
\end{equation}
where $\left[  x\right]  $ and $\left\{  x\right\}  $ are the integer and
fractional parts of $x$, respectively. Given that $f_{H}$ and $nc_{2}$ can now
be represented as $f_{H}=m\left(  f_{H}\right)  +\theta\left(  f_{H}\right)  $
and $nc_{2}=$sgn~$n\left[  m\left(  \left\vert n\right\vert c_{2}\right)
+\theta\left(  \left\vert n\right\vert c_{2}\right)  \right]  $, respectively,
the result of the minimization is%
\begin{gather}
G_{em}\left(  \left\vert n\right\vert ;f_{H}\right)  =\min_{n_{1}%
,\text{sgn}~n}G_{em}\left(  n_{1},n;f_{H}\right) \nonumber\\
=G_{em}\left(  -m\left(  f_{H}\right)  +m\left(  \left\vert n\right\vert
c_{2}\right)  \text{sgn}~\theta\left(  f_{H}\right)  \text{sgn}~\theta\left(
\left\vert n\right\vert c_{2}\right)  ,\left\vert n\right\vert \text{sgn}%
~\theta\left(  f_{H}\right)  \text{sgn}~\theta\left(  \left\vert n\right\vert
c_{2}\right)  ;f_{H}\right) \nonumber\\
=\frac{\Phi_{0}^{2}\varepsilon}{2\mathcal{L}_{m}}\left[  \left\vert
\theta\left(  f_{H}\right)  \right\vert -\left\vert \theta\left(  \left\vert
n\right\vert c_{2}\right)  \right\vert \right]  ^{2}\left(  1-\varepsilon
\right)  . \label{2.37}%
\end{gather}

\section{Soliton states}

\subsection{Soliton self-energy}

The variation of (\ref{2.15}) with respect to $\phi$, under the condition of
single-valuedness of variations $\delta\phi$, yields a static two-dimensional
sine-Gordon equation in polar coordinates,%
\begin{equation}
\frac{1}{r^{2}}\frac{\partial^{2}\phi}{\partial\varphi^{2}}+\frac{1}{r}%
\frac{\partial}{\partial r}\left(  r\frac{\partial\phi}{\partial r}\right)
=\frac{\text{sgn}~\gamma}{l^{2}}\sin\phi,\quad r\in\left(  r_{1},r_{2}\right)
,\quad\varphi\in\left(  0,2\pi\right)  , \label{3.1}%
\end{equation}
and the boundary conditions%
\begin{equation}
\left.  \frac{1}{r}\frac{\partial\phi}{\partial\varphi}\right\vert
_{\varphi=0}=\left.  \frac{1}{r}\frac{\partial\phi}{\partial\varphi
}\right\vert _{\varphi=2\pi};\quad\left.  \frac{\partial\phi}{\partial
r}\right\vert _{r=r_{1}}=\left.  \frac{\partial\phi}{\partial r}\right\vert
_{r=r_{2}}=0. \label{3.2}%
\end{equation}
[These boundary conditions should, of course, be complemented by a condition
on $\phi$ resulting from (\ref{1.8}).]

However, equation (\ref{3.1}), in its exact form, by far exceeds the accuracy
of our calculations in the previous section [see expressions (\ref{2.29}),
(\ref{2.30}) and relations (\ref{a8})]. Discarding in (\ref{3.1}) terms of
order $\frac{d}{R}$ and $\frac{d^{2}}{R^{2}}$, we arrive at a two-dimensional
sine-Gordon equation in "Cartesian coordinates":%
\begin{equation}
\frac{\partial^{2}\phi}{\partial\varphi^{2}}+\frac{\partial^{2}\phi}%
{\partial\rho^{2}}=\frac{R^{2}\text{sgn}~\gamma}{l^{2}}\sin\phi,\quad
\rho\equiv\frac{r}{R}. \label{3.3}%
\end{equation}
Solutions to (\ref{3.3}), minimizing the functional (\ref{2.15}), should not
depend on $\rho$ for symmetry reasons [which, of course, is compatible with
boundary conditions (\ref{3.2})].

Thus, the phase $\phi=\phi\left(  \varphi\right)  $ satisfies the following
boundary-value problem:%
\begin{gather}
\frac{d^{2}\phi}{d\varphi^{2}}=\frac{R^{2}\text{sgn}~\gamma}{l^{2}}\sin
\phi,\quad\varphi\in\left(  0,2\pi\right)  ;\nonumber\\
\phi\left(  2\pi\right)  =\phi\left(  0\right)  +2\pi n\quad\left(  n=\pm
1,\pm2,\ldots\right)  ,\quad\frac{d\phi}{d\varphi}\left(  2\pi\right)
=\frac{d\phi}{d\varphi}\left(  0\right)  . \label{3.4}%
\end{gather}
The solution of (\ref{3.4}) is straightforward:\cite{KG06}%
\begin{equation}
\phi_{n}\left(  \varphi\right)  =\frac{\left(  1+\text{sgn}~\gamma\right)\pi
}{2}+2\text{am}~\left(  \frac{nK\left(  k_{n}\right)  }{\pi}\left(
\varphi-\varphi_{n0}\right)  ,k_{n}\right)  , \label{3.5}%
\end{equation}
where am~$u$ is the elliptic amplitude,\cite{AS65} $K\left(  k\right)  $ is
the complete elliptic integral of the first kind,\cite{AS65} $\varphi_{n0}$
are arbitrary constants, and $k_{n}$ ($n=\pm1,\pm2,\ldots$) satisfy the
equations%
\begin{equation}
\left\vert n\right\vert k_{n}K\left(  k_{n}\right)  =\frac{\pi R}{l},\quad
n=\pm1,\pm2,\ldots. \label{3.6}%
\end{equation}

Particular examples of solutions (\ref{3.5}) that possess asymptotics in terms
of elementary functions are relegated to Appendix C. Nonetheless, the very
special class of \textit{exact} elementary solutions is worth being presented
here: namely, the \textit{non}-soliton nontrivial topological solutions
corresponding to the physical case of the absence of interband coupling
($\left\vert \gamma\right\vert =0$). These solutions can be obtained from
(\ref{3.5}) by the limit procedure%
\begin{equation}
k_{n}\rightarrow0,\quad n=\pm1,\pm2,\ldots, \label{3.7}%
\end{equation}
and they have the general form%
\begin{equation}
\phi_{n}\left(  \varphi\right)  =n\varphi+\varphi_{0}. \label{3.8}%
\end{equation}
They are necessarily minimizers of (\ref{2.15}) [i.e., at these solutions
$\delta^{2}F_{sol}>0$, because the functional (\ref{2.15}) is quadratic in the
case $\left\vert \gamma\right\vert =0$], and their self-energy is%
\begin{equation}
F_{sol}^{\left(  0\right)  }\left(  n\right)  =\lim_{k_{n}\rightarrow0}%
F_{sol}\left(  n\right)  =\frac{\Phi_{0}^{2}\varepsilon}{2\mathcal{L}_{m}%
}\left\vert n\right\vert ^{2}c_{1}c_{2}. \label{3.9}%
\end{equation}

If $\left\vert \gamma\right\vert \neq0$, the functional (\ref{2.15}) is
non-quadratic, and we should analyze the second variation of (\ref{2.15}) in
more detail. To this end,\cite{KG06,KG07} we turn to the Sturm-Liouville
problem%
\begin{gather}
-\frac{d^{2}\psi}{d\varphi^{2}}+\cos\not \phi _{n}\psi=\mu\psi,\quad\varphi
\in\left(  0,2\pi\right)  ;\label{3.10}\\
\psi\left(  0\right)  =\psi\left(  2\pi\right)  ,\quad\frac{d\psi}{d\varphi
}\left(  0\right)  =\frac{d\psi}{d\varphi}\left(  2\pi\right)  ,\nonumber
\end{gather}
where $\not \phi _{n}$ is a given solution from the set (\ref{3.5}). As shown
in Refs. \cite{KG06,KG07},%
\[
\left.  \delta^{2}F_{sol}\right\vert _{\phi=\phi_{n}}\geq\mu_{0}\int_{0}%
^{2\pi}\left\vert \delta\phi_{n}\right\vert ^{2}d\varphi,
\]
where $\mu_{0}$ is lowest eigenvalue of the problem (\ref{3.10}). In our case,
both $\mu_{0}$ and the corresponding eigenfunction $\psi_{0}$ can be readily
found:%
\[
\mu_{0}=0,\quad\psi_{0}=\text{const~dn}\left(  \frac{nK\left(  k_{n}\right)
}{\pi}\left(  \varphi-\varphi_{n0}\right)  ,k_{n}\right)  ,\text{~}%
\]
where dn$\,u=\frac{d\,\text{am}\,u}{du}$.\cite{AS65} This means that $\left.
\delta^{2}F_{sol}\right\vert _{\phi=\phi_{n}}\geq0$, and soliton states turn
out to be \textit{indifferently} stable states. Indeed, the zero value of
$\mu_{0}$ should be attributed to the existence of a zero-frequency
"rotational mode" (by analogy with the well-known\cite{Ja77} translational
mode in quantum field theories) that restores rotational symmetry broken by
the formation of solitons. To prove this, consider a small variation of
$\phi_{n}$ induced by a small variation of the constant of integration
$\varphi_{n0}$:%
\[
\phi_{n}\left(  \varphi\right)  \rightarrow\phi_{n}\left(  \varphi+\frac{\pi
}{nK}\alpha\right)  \approx\phi_{n}\left(  \varphi\right)  +\alpha
\text{dn}\left(  \frac{nK\left(  k_{n}\right)  }{\pi}\left(  \varphi
-\varphi_{n0}\right)  ,k_{n}\right)  ,\quad\left\vert \alpha\right\vert \ll1.
\]
From the above, we see that $\delta\phi_{n}\varpropto\psi_{0}$.

Now that local stability of soliton solutions is established, we proceed with
a discussion of soliton self-energy. It is obtained by the substitution of
solutions (\ref{3.5}) into (\ref{2.15}) and has the form%
\begin{equation}
F_{sol}\left(  n\right)  =\frac{\Phi_{0}^{2}\varepsilon}{\mathcal{L}_{m}}%
\frac{2\left\vert n\right\vert ^{2}}{\pi^{2}}c_{1}c_{2}K\left(  k_{n}\right)
\left[  2E\left(  k_{n}\right)  -\left(  1-k_{n}^{2}\right)  K\left(
k_{n}\right)  \right]  , \label{3.12}%
\end{equation}
where $E\left(  k\right)  $ is the complete elliptic integral of the second
kind.\cite{AS65}

First, we note that the constants of integration $\varphi_{n0}$ that figure in
(\ref{3.5}) drop out of the right-hand side of (\ref{3.12}), as they should.
The self-energy does not depend on the sign of $\gamma$ and of $n$, either. By
considering (formally) $\left\vert n\right\vert $ as a continuous variable, we
get%
\[
\frac{\partial F_{sol}\left(  n\right)  }{\partial\left\vert n\right\vert
}=\frac{\Phi_{0}^{2}\varepsilon}{\mathcal{L}_{m}}\frac{4\left\vert
n\right\vert R}{\pi l}c_{1}c_{2}E\left(  k_{n}\right)  >0,
\]
which means that $F_{sol}\left(  n\right)  $ increases monotonically with an
increase in $\left\vert n\right\vert $, as could be expected. However, in
contrast to the case $\left\vert \gamma\right\vert =0$ [see (\ref{3.9})], the
growth of $F_{sol}\left(  n\right)  $ is slower than $\left\vert n\right\vert
^{2}$, because%
\[
\frac{\partial}{\partial\left\vert n\right\vert }\left[  \frac{F_{sol}\left(
n\right)  }{\left\vert n\right\vert ^{2}}\right]  =-\frac{\Phi_{0}%
^{2}\varepsilon}{\mathcal{L}_{m}}\frac{4R}{\pi\left\vert n\right\vert ^{2}%
l}c_{1}c_{2}\frac{E\left(  k_{n}\right)  -\left(  1-k_{n}^{2}\right)  K\left(
k_{n}\right)  }{k_{n}}<0.
\]

Given that%
\[
\frac{\partial F_{sol}\left(  n\right)  }{\partial k_{n}}=\frac{\Phi_{0}%
^{2}\varepsilon}{\mathcal{L}_{m}}\frac{4\left\vert n\right\vert ^{2}}{\pi^{2}%
}c_{1}c_{2}\frac{E\left(  k_{n}\right)  \left[  E\left(  k_{n}\right)
-\left(  1-k_{n}^{2}\right)  K\left(  k_{n}\right)  \right]  }{k_{n}\left(
1-k_{n}^{2}\right)  }>0,\quad k_{n}\in\left(  0,1\right)  ,
\]
the self-energy increases monotonically with an increase in $k_{n}$ on the
whole interval $\left(  0,1\right)  $. The minimal value of (\ref{3.12}) is
achieved at $k_{n}=0$ and is given by (\ref{3.9}). In view of the relation%
\[
\frac{\partial k_{n}}{\partial\left(  \frac{l}{R}\right)  }=-\frac{\pi R^{2}%
}{\left\vert n\right\vert ^{2}l^{2}}\frac{1-k_{n}^{2}}{E\left(  k_{n}\right)
}<0,
\]
the self-energy decreases monotonically with an increase in $\frac{l}{R}%
\in\left(  0,\infty\right)  $ (for a given $\left\vert n\right\vert $). [In
other words, $F_{sol}$ is an \textit{increasing} function of the interband
coupling parameter $\left\vert \gamma\right\vert $: see the definition of $l $
in (\ref{2.15}).]

\subsection{Thermodynamic metastability}

According to (\ref{2.12}), (\ref{2.37}) and (\ref{3.12}) the minimal Gibbs
free energy of soliton states with a given $\left\vert n\right\vert $ in the
field $H$ can be represented as follows:%
\begin{gather}
G\left(  \left\vert n\right\vert ;f_{H}\right)  =F_{0}+\frac{\Phi_{0}%
^{2}\varepsilon}{2\mathcal{L}_{m}}\left[  \left[  \left\vert \theta\left(
f_{H}\right)  \right\vert -\left\vert \theta\left(  \left\vert n\right\vert
c_{2}\right)  \right\vert \right]  ^{2}\left(  1-\varepsilon\right)  \right.
\nonumber\\
\left.  +\frac{4\left\vert n\right\vert ^{2}}{\pi^{2}}c_{1}c_{2}K\left(
k_{n}\right)  \left[  2E\left(  k_{n}\right)  -\left(  1-k_{n}^{2}\right)
K\left(  k_{n}\right)  \right]  \right]  . \label{3.14}%
\end{gather}
To analyze thermodynamic stability of soliton solutions, we should compare
expression (\ref{3.14}) for $\left\vert n\right\vert \geq1$ with the Gibbs
free energy of the states with $\left\vert n\right\vert =0$:%
\begin{equation}
G\left(  0;f_{H}\right)  =F_{0}+\frac{\Phi_{0}^{2}\varepsilon}{2\mathcal{L}%
_{m}}\left\vert \theta\left(  f_{H}\right)  \right\vert ^{2}\left(
1-\varepsilon\right)  . \label{3.15}%
\end{equation}

With this in mind, we first note that, for $\left\vert n\right\vert \geq1$,
the energy $G\left(  \left\vert n\right\vert ;f_{H}\right)  $ increases
monotonically with an increase in $\left\vert n\right\vert $: see Appendix B
for a proof. [Contrary to what may seem, this fact is by no means obvious,
because the electromagnetic term in (\ref{3.14}) may \textit{decrease} with an
increase in $\left\vert n\right\vert $.] Furthermore, since expression
(\ref{3.9}) provides the greatest lower bound for soliton self-energies, we
can restrict ourselves to the case $\left\vert n\right\vert =1$ and $k_{n}=0$.
Bearing in mind that $c_{2}\in\left(  0,\frac{1}{2}\right]  $ by assumption
(see the end of Section III), we arrive at the following important
inequalities:%
\begin{gather}
\Delta G\left(  1;f_{H}\right)  \equiv G\left(  1;f_{H}\right)  -G\left(
0;f_{H}\right) \nonumber\\
\geq\Delta G^{\left(  0\right)  }\left(  1;f_{H}\right)  \equiv\lim
_{k_{n}\rightarrow0}G\left(  1;f_{H}\right)  -G\left(  0;f_{H}\right)
\nonumber\\
=\frac{\Phi_{0}^{2}\varepsilon}{2\mathcal{L}_{m}}c_{2}\left[  1-c_{2}%
\varepsilon-2\left\vert \theta\left(  f_{H}\right)  \right\vert \left(
1-\varepsilon\right)  \right]  >0. \label{3.16}%
\end{gather}

The above inequalities clearly demonstrate thermodynamic metastability of
soliton states and bring to light certain subtle physical points. In
particular,%
\begin{equation}
\max_{f_{H}}\Delta G\left(  1;f_{H}\right)  =\left.  \Delta G\left(
1;f_{H}\right)  \right\vert _{f_{H}=p}\geq\frac{\Phi_{0}^{2}\varepsilon
}{2\mathcal{L}_{m}}c_{2}\left(  1-c_{2}\varepsilon\right)  ,\quad
p=0,1,2,\ldots. \label{3.17}%
\end{equation}
In contrast,%
\begin{equation}
\min_{f_{H}}\Delta G\left(  1;f_{H}\right)  =\left.  \Delta G\left(
1;f_{H}\right)  \right\vert _{f_{H}=p+\frac{1}{2}}\geq\frac{\Phi_{0}%
^{2}\varepsilon^{2}}{2\mathcal{L}_{m}}c_{1}c_{2},\quad p=0,1,2,\ldots,
\label{3.18}%
\end{equation}
which shows that $\min_{f_{H}}\Delta G\left(  1;f_{H}\right)  $ can be much
smaller than the magnetic Gibbs free energy of the zero-soliton states [see
(\ref{2.33})], provided that $0<c_{2}\ll c_{1}<1$ [i.e., when $\lambda_{1}%
\ll\lambda_{2}<\infty$, see (\ref{2.11})] and $\frac{l}{R}\gg1$ (weak
interband coupling).

\begin{figure}[ptb]
\begin{center}
$%
\begin{array}
[c]{cc}%
\includegraphics[width=0.5\textwidth]{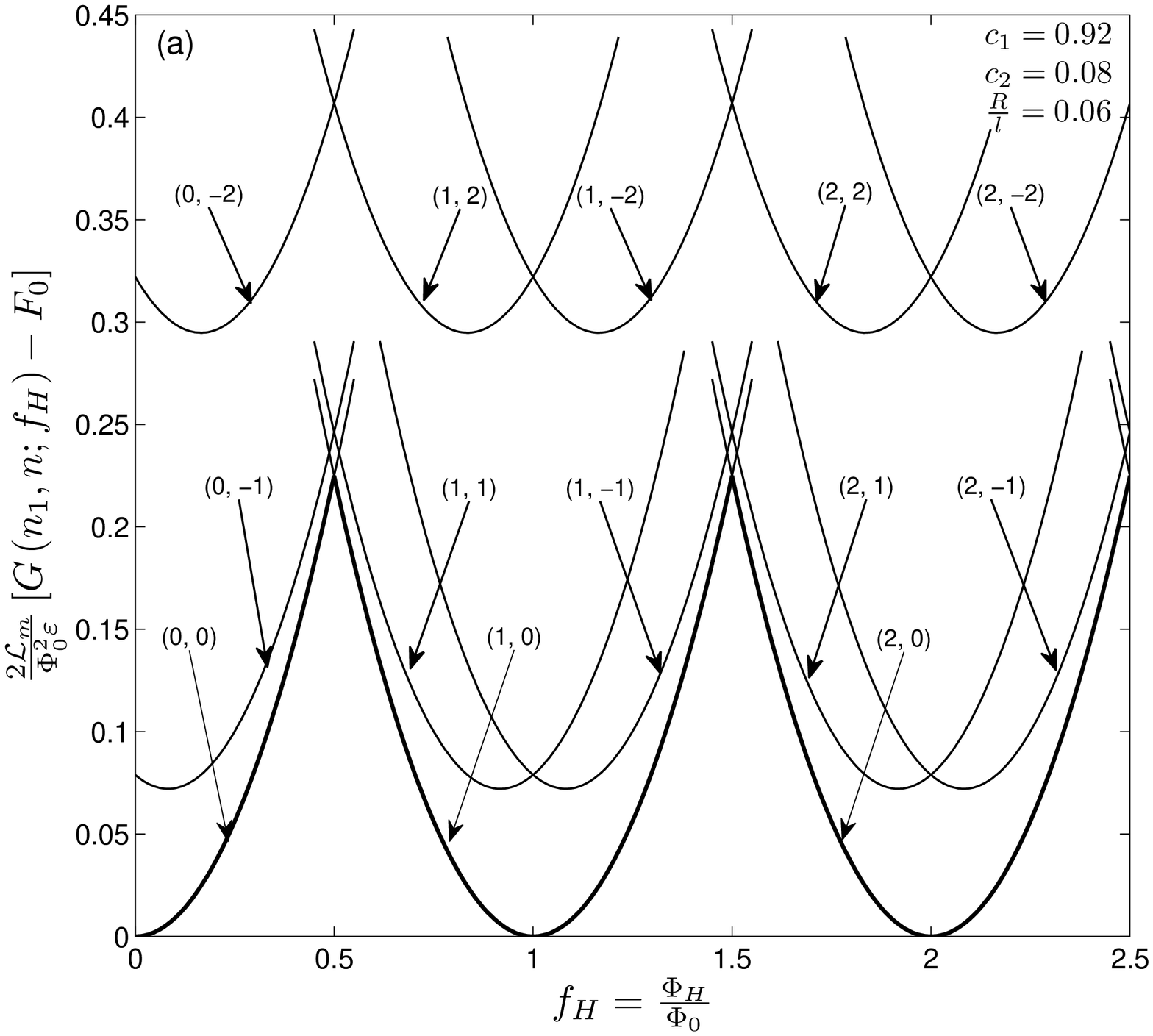} &
\includegraphics[width=0.5\textwidth]{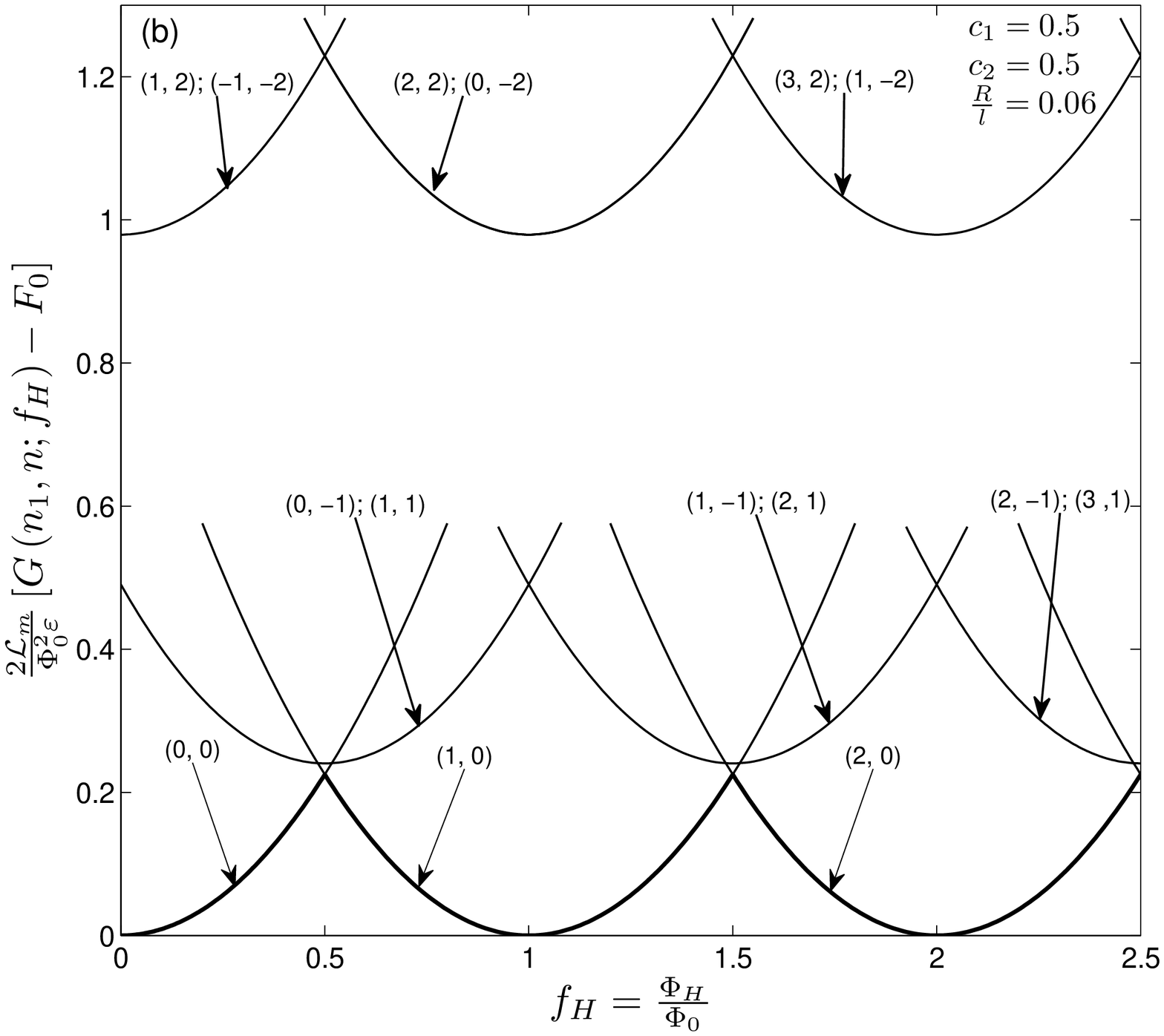}\\
\includegraphics[width=0.5\textwidth]{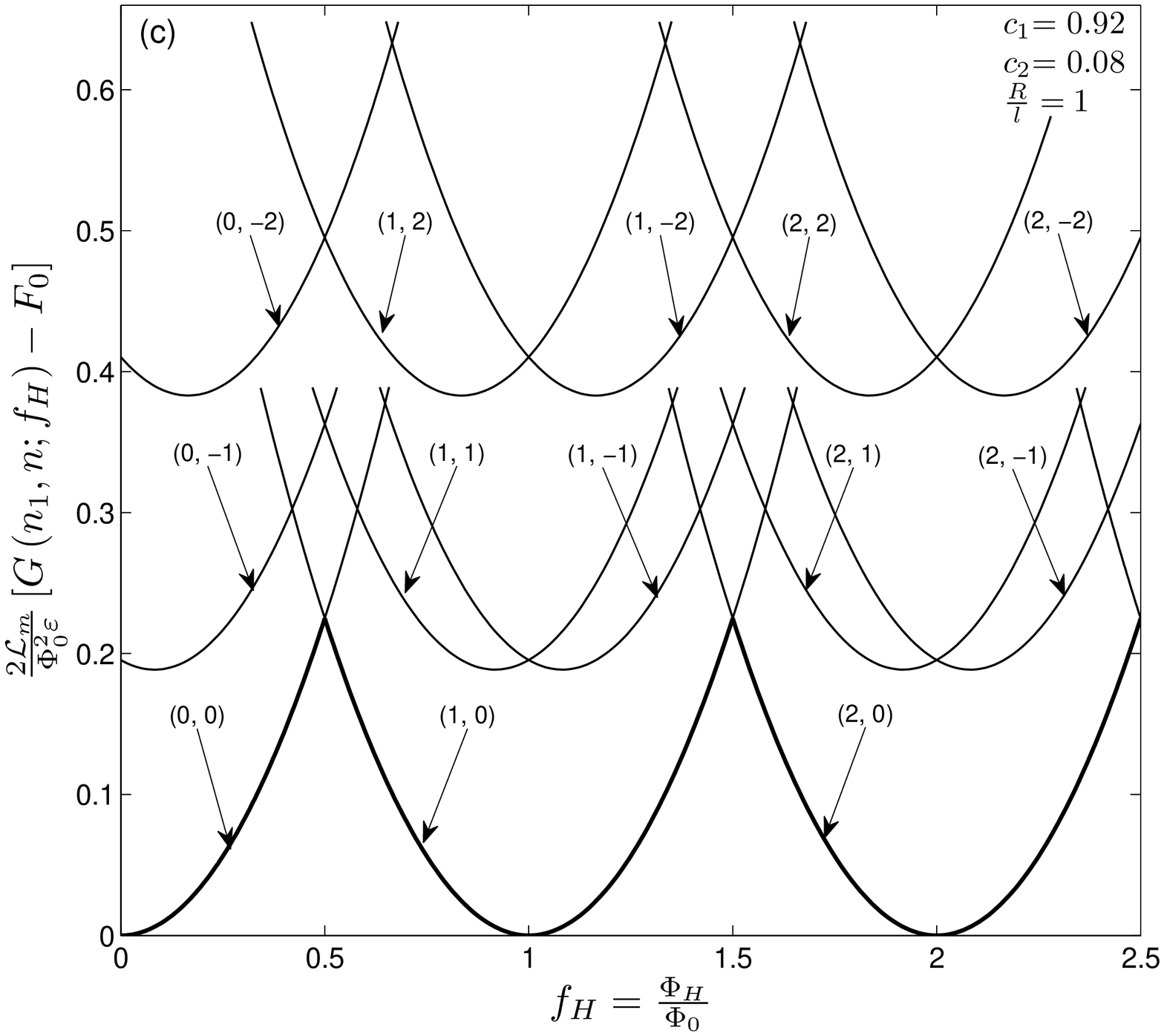} &
\includegraphics[width=0.5\textwidth]{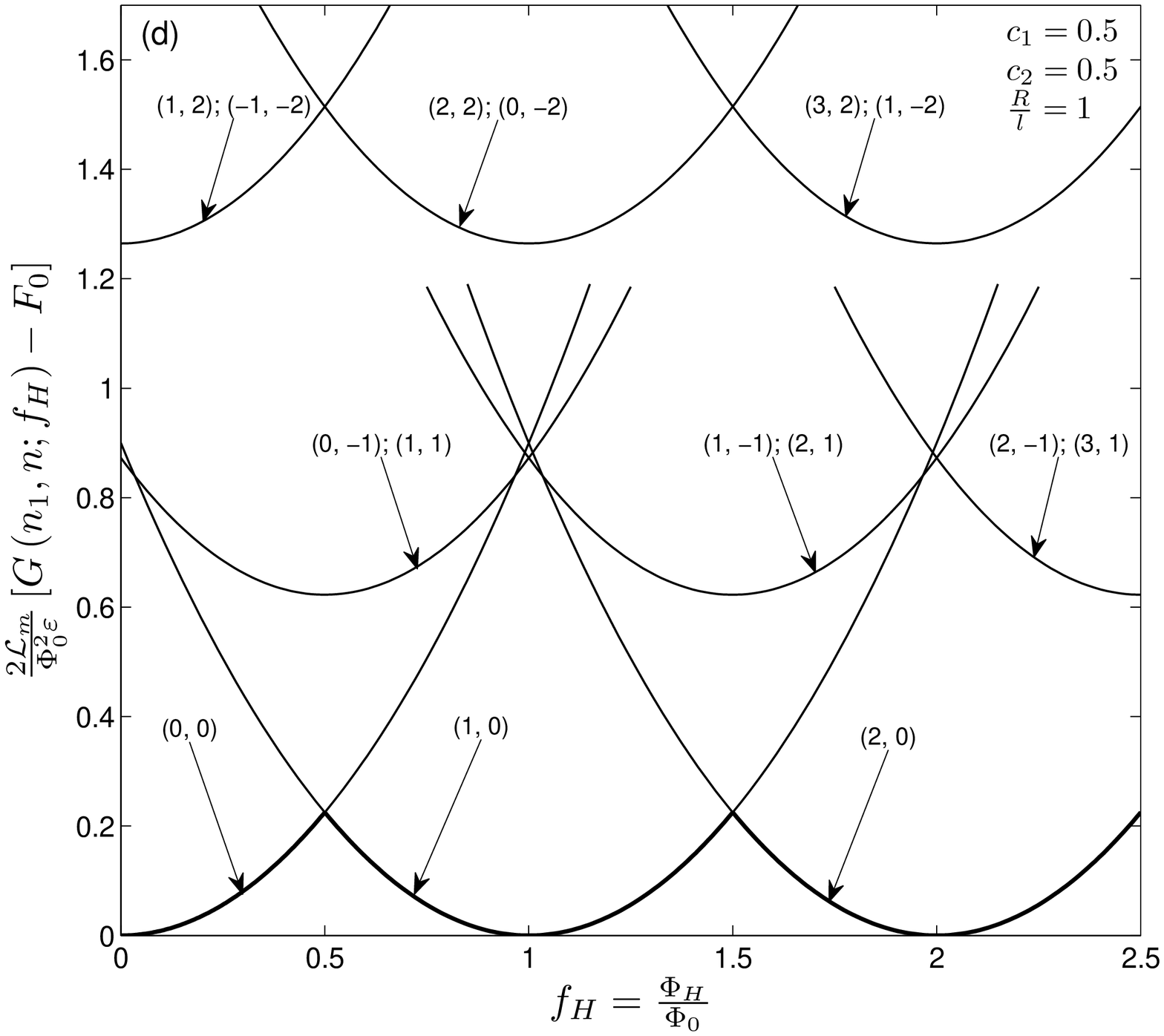}
\end{array}
$
\end{center}
\caption{Gibbs free energy of several different topological states $\left(
n_{1},n\right)  $ ($\left\vert n\right\vert =0,1,2$) for typical values of the
parameters $c_{1}$, $c_{2}$, and $\frac{R}{l}$. Note double degeneracy of
soliton states in the case $c_{1}=c_{2}=\frac{1}{2}$ and see the text for
further explanations.}%
\label{fig:2}%
\end{figure}

In Fig. \ref{fig:2}, we plot the Gibbs free energy of several different
topological states $\left(  n_{1},n\right)  $ ($\left\vert n\right\vert
=0,1,2$). Thermodynamically stable zero-soliton states are denoted by thick
solid lines. Minima of the Gibbs free energy of soliton states represent the
soliton self-energy and occur when the self-induced flux $\Phi_{i}$
compensates for the external flux $\Phi_{H}$, i.e., when%
\[
f_{H}+n_{1}-nc_{2}=0,
\]
[see relations (\ref{2.31}), (\ref{2.34}) and (\ref{2.37})]. In the very
special case, when $c_{2}=c_{1}=\frac{1}{2}$ and $n=2n_{1}$, no flux is
induced ($\Phi_{i}=0$), and minima of the soliton Gibbs free energy occur at
$H=0$.

\section{Summary and conclusions}

Summarizing, we have presented (in the framework of the Ginzburg-Landau
approach) a self-consistent theory of specific soliton states that constitute
a distinctive feature of two-band superconductivity in mesoscopic
multiply-connected samples. Although our mathematical consideration concerns
the concrete geometry of Fig. 1, the final results can be expressed in terms
of the magnetic and kinetic inductance (see Ref. \cite{LL68}) and, therefore,
should apply to a much wider class of structures. This allows us to make
several generalizing remarks.

As the predicted fractional magnetic vortices in bulk two-band
superconductors,\cite{B02} the soliton states considered here prove to be
thermodynamically metastable. However, the minimal energy gap between the
lowest-lying single-soliton states and thermodynamically stable zero-soliton
states can be much smaller than the magnetic Gibbs free energy of the latter
states, provided that the intraband "penetration depths" (\ref{2.9}) differ
substantially and the interband coupling is weak. (In order to establish this
important physical fact, we had to evaluate self-consistantly the vector
potential. The results of this evaluation may be of interest in themselves.)

Our consideration encompasses in a natural way the case of superconducting
Josephson-coupled bilayer structures studied experimentally in Ref.
\cite{BKHM06}. Our conclusion that the self-energy of soliton states increases
monotonically with an increase in the strength of interband coupling
qualitatively agrees with the observations reported therein.

Furthermore, as a particular limit, our consideration contains the case of
zero interband coupling. In view of the recently discussed possibility of
independent superconductivity of electrons and protons in a liquid metallic
state of hydrogen,\cite{BSA04} some of our results may find application in
this situation as well.

Finally, the exact soliton solutions derived in this paper should be compared
with the exact soliton solutions representing equilibrium Josephson vortices
in a superconducting tunnel junction.\cite{KG06,KG07} In particular, Josephson
vortices are pinned in their equilibrium positions owing to interaction with
the edges of the junction. In contrast, in rotationally symmetric
doubly-connected two-band superconductors, \ soliton positions are not fixed,
which gives rise to a specific zero-frequency rotational mode. However, any
defects that break rotational symmetry must cause soliton pinning. The effect
of this pinning requires a separate discussion. In conclusion, we hope that
our paper will stimulate further experimental and theoretical studies of the
intriguing phenomenon of soliton states in two-band superconductors.

\appendix

\section{Exact solution of the boundary-value problem for the vector
potential}

The exact solution to (\ref{2.28}) has the following form:%
\begin{gather}
A\left(  r\right)  =\frac{rh_{O}}{2},\quad r\in\left[  0,r_{1}\right]
;\nonumber\\
=-\frac{\Phi_{0}}{2\pi r}q\left(  n_{1},n_{2}\right)  +C_{1}I_{1}\left(
\frac{r}{\lambda}\right)  +C_{2}K_{1}\left(  \frac{r}{\lambda}\right)  ,\quad
r\in\left(  r_{1},r_{2}\right]  ; \label{a1}%
\end{gather}%
\begin{gather}
h\left(  r\right)  =h_{O},\quad r\in\left[  0,r_{1}\right]  ;\nonumber\\
=\frac{C_{1}}{\lambda}I_{0}\left(  \frac{r}{\lambda}\right)  -\frac{C_{2}%
}{\lambda}K_{0}\left(  \frac{r}{\lambda}\right)  ,\quad r\in\left(
r_{1},r_{2}\right]  ; \label{a2}%
\end{gather}%
\begin{equation}
h_{O}=\frac{2\lambda Hf_{2}\left(  \frac{r_{1}}{\lambda},\frac{r_{1}}{\lambda
}\right)  }{r_{1}f_{1}\left(  \frac{r_{2}}{\lambda},\frac{r_{1}}{\lambda
}\right)  +2\lambda f_{2}\left(  \frac{r_{2}}{\lambda},\frac{r_{1}}{\lambda
}\right)  }-\frac{\Phi_{0}q\left(  n_{1},n_{2}\right)  f_{1}\left(
\frac{r_{2}}{\lambda},\frac{r_{1}}{\lambda}\right)  }{\pi r_{1}\left[
r_{1}f_{1}\left(  \frac{r_{2}}{\lambda},\frac{r_{1}}{\lambda}\right)
+2\lambda f_{2}\left(  \frac{r_{2}}{\lambda},\frac{r_{1}}{\lambda}\right)
\right]  }; \label{a3}%
\end{equation}%
\[
C_{1}\equiv\frac{\lambda H\left[  r_{1}K_{0}\left(  \frac{r_{1}}{\lambda
}\right)  +2\lambda K_{1}\left(  \frac{r_{1}}{\lambda}\right)  \right]
}{r_{1}f_{1}\left(  \frac{r_{2}}{\lambda},\frac{r_{1}}{\lambda}\right)
+2\lambda f_{2}\left(  \frac{r_{2}}{\lambda},\frac{r_{1}}{\lambda}\right)
}+\frac{\lambda\Phi_{0}q\left(  n_{1},n_{2}\right)  K_{0}\left(  \frac{r_{2}%
}{\lambda}\right)  }{\pi r_{1}\left[  r_{1}f_{1}\left(  \frac{r_{2}}{\lambda
},\frac{r_{1}}{\lambda}\right)  +2\lambda f_{2}\left(  \frac{r_{2}}{\lambda
},\frac{r_{1}}{\lambda}\right)  \right]  },
\]%
\[
C_{2}\equiv\frac{\lambda H\left[  r_{1}I_{0}\left(  \frac{r_{1}}{\lambda
}\right)  -2\lambda I_{1}\left(  \frac{r_{1}}{\lambda}\right)  \right]
}{r_{1}f_{1}\left(  \frac{r_{2}}{\lambda},\frac{r_{1}}{\lambda}\right)
+2\lambda f_{2}\left(  \frac{r_{2}}{\lambda},\frac{r_{1}}{\lambda}\right)
}+\frac{\lambda\Phi_{0}q\left(  n_{1},n_{2}\right)  I_{0}\left(  \frac{r_{2}%
}{\lambda}\right)  }{\pi r_{1}\left[  r_{1}f_{1}\left(  \frac{r_{2}}{\lambda
},\frac{r_{1}}{\lambda}\right)  +2\lambda f_{2}\left(  \frac{r_{2}}{\lambda
},\frac{r_{1}}{\lambda}\right)  \right]  };
\]%
\[
f_{1}\left(  x,y\right)  \equiv I_{0}\left(  x\right)  K_{0}\left(  y\right)
-I_{0}\left(  y\right)  K_{0}\left(  x\right)  ,\quad f_{2}\left(  x,y\right)
\equiv I_{0}\left(  x\right)  K_{1}\left(  y\right)  +I_{1}\left(  y\right)
K_{0}\left(  x\right)  .\,
\]
Here, $I_{\nu}\left(  x\right)  $ and $K_{\nu}\left(  x\right)  $ are modified
Bessel functions of order $\nu=0,1$;\cite{AS65} $h_{O}$ is the constant
magnetic field in the opening. Expressions (\ref{a1})-(\ref{a3}) are greatly
simplified under condition (\ref{1.5}):%
\begin{gather}
A\left(  r\right)  =\frac{rh_{O}}{2},\quad r\in\left[  0,r_{1}\right]
;\nonumber\\
=-\frac{\Phi_{0}q\left(  n_{1},n_{2}\right)  }{2\pi r}\frac{\sinh\frac
{d}{\lambda}+\frac{2\lambda}{r_{1}}\left(  \cosh\frac{d}{\lambda}-\sqrt
{\frac{r}{r_{1}}}\cosh\frac{r_{2}-r}{\lambda}\right)  }{\sinh\frac{d}{\lambda
}+\frac{2\lambda}{r_{1}}\cosh\frac{d}{\lambda}}\nonumber\\
+\lambda H\sqrt{\frac{r_{2}}{r}}\frac{\cosh\frac{r-r_{1}}{\lambda}%
+\frac{2\lambda}{r_{1}}\sinh\frac{r-r_{1}}{\lambda}}{\sinh\frac{d}{\lambda
}+\frac{2\lambda}{r_{1}}\cosh\frac{d}{\lambda}},\quad r\in\left(  r_{1}%
,r_{2}\right]  ; \label{a4}%
\end{gather}%
\begin{gather}
h\left(  r\right)  =h_{O},\quad r\in\left[  0,r_{1}\right]  ;\nonumber\\
=-\frac{\Phi_{0}q\left(  n_{1},n_{2}\right)  }{\pi r_{1}\sqrt{rr_{1}}}%
\frac{\sinh\frac{r_{2}-r}{\lambda}}{\sinh\frac{d}{\lambda}+\frac{2\lambda
}{r_{1}}\cosh\frac{d}{\lambda}}\nonumber\\
+H\sqrt{\frac{r_{2}}{r}}\frac{\sinh\frac{r-r_{1}}{\lambda}+\frac{2\lambda
}{r_{1}}\cosh\frac{r-r_{1}}{\lambda}}{\sinh\frac{d}{\lambda}+\frac{2\lambda
}{r_{1}}\cosh\frac{d}{\lambda}},\quad r\in\left(  r_{1},r_{2}\right]  ;
\label{a5}%
\end{gather}%
\begin{equation}
h_{O}=\frac{-\frac{\Phi_{0}q\left(  n_{1},n_{2}\right)  }{\pi r_{1}^{2}}%
\sinh\frac{d}{\lambda}+\frac{2\lambda}{r_{1}}\sqrt{\frac{r_{2}}{r_{1}}}%
H}{\sinh\frac{d}{\lambda}+\frac{2\lambda}{r_{1}}\cosh\frac{d}{\lambda}}.
\label{a6}%
\end{equation}

Expressions (\ref{a4})-(\ref{a6}) should be compared with analogous
expressions for a single-band-superconducting cylinder.\cite{G62} As can be
easily seen, the cylinder exhibits a considerable Meissner effect under the
conditions%
\begin{equation}
\frac{d}{\lambda}\ll1,\quad\frac{dR}{2\lambda^{2}}\gg1. \label{a7}%
\end{equation}

In contrast, in the opposite case, when condition (\ref{1.6}) is fulfilled,
the Meissner effect is small, and expressions (\ref{a4})-(\ref{a6}) can be
readily expanded up to first-order terms in $\varepsilon$. Taking into account
a hierarchy of the small parameters of the problem,%
\begin{gather}
\frac{d}{r_{1}}\approx\frac{d}{r_{2}}\approx\frac{d}{R}\equiv2\varepsilon
\frac{\lambda^{2}}{R^{2}},\quad\frac{d}{\lambda}\equiv2\varepsilon
\frac{\lambda}{R};\nonumber\\
\frac{d}{R}\ll\frac{d}{\lambda}\ll\varepsilon,\frac{\lambda}{R}\ll1,
\label{a8}%
\end{gather}
we arrive at the first-order expressions (\ref{2.29}) and (\ref{2.30}).

\section{A proof of the inequality $\left.  G\left(  \left\vert n\right\vert
;f_{H}\right)  \right\vert _{\left\vert n\right\vert >1}>G\left(
1;f_{H}\right)  $}

Consider the expression%
\begin{gather}
\frac{2\mathcal{L}_{m}}{\Phi_{0}^{2}\varepsilon}\Delta G\left(  \left\vert
n\right\vert ;f_{H}\right)  \equiv\frac{2\mathcal{L}_{m}}{\Phi_{0}%
^{2}\varepsilon}\left[  G\left(  \left\vert n\right\vert ;f_{H}\right)
-G\left(  0;f_{H}\right)  \right] \nonumber\\
=-\left\vert \theta\left(  \left\vert n\right\vert c_{2}\right)  \right\vert
\left[  2\left\vert \theta\left(  f_{H}\right)  \right\vert -\left\vert
\theta\left(  \left\vert n\right\vert c_{2}\right)  \right\vert \right]
\left(  1-\varepsilon\right)  +\frac{4\left\vert n\right\vert ^{2}}{\pi^{2}%
}c_{1}c_{2}K\left(  k_{n}\right)  \left[  2E\left(  k_{n}\right)  -\left(
1-k_{n}^{2}\right)  K\left(  k_{n}\right)  \right]  \label{b1}%
\end{gather}
that follows directly from (\ref{3.14}). Our task is to prove that the
right-hand side of (\ref{b1}) for $\left\vert n\right\vert >1$ is larger than
for $\left\vert n\right\vert =1$. Given that $c_{2}\in\left(  0,\frac{1}%
{2}\right]  $ by assumption (see the end of Section III), it is sufficient to
provide a proof for $1<\left\vert n\right\vert \leq\left[  \frac{1}{2c_{2}%
}\right]  +1$, where $\left[  \frac{1}{2c_{2}}\right]  $ is the integer part
of $\frac{1}{2c_{2}}$. Indeed, the first term on the right-hand side of
(\ref{b1}) satisfies the inequality%
\[
\left\vert \left\vert \theta\left(  \left\vert n\right\vert c_{2}\right)
\right\vert \left[  2\left\vert \theta\left(  f_{H}\right)  \right\vert
-\left\vert \theta\left(  \left\vert n\right\vert c_{2}\right)  \right\vert
\right]  \left(  1-\varepsilon\right)  \right\vert \leq\frac{1}{4}\left(
1-\varepsilon\right)  ,
\]
whereas%
\[
\frac{4K\left(  k_{n}\right)  }{\pi^{2}}\left[  2E\left(  k_{n}\right)
-\left(  1-k_{n}^{2}\right)  K\left(  k_{n}\right)  \right]  \geq1.
\]
Therefore, for $\left\vert n\right\vert >\left[  \frac{1}{2c_{2}}\right]  +1$,
when $\left\vert n\right\vert c_{2}>\frac{1}{2}$, any possible decrease in the
first term on the right-hand side of (\ref{b1}) due to an increase in
$\left\vert n\right\vert $ cannot compensate for an incurred increase in the
second term.

For $1<\left\vert n\right\vert <\left[  \frac{1}{2c_{2}}\right]  +1$, there
holds the relation $\left\vert n\right\vert c_{2}\leq\frac{1}{2}$, and
(\ref{b1}) becomes%
\begin{gather*}
\frac{2\mathcal{L}_{m}}{\Phi_{0}^{2}\varepsilon}\left.  \Delta G\left(
\left\vert n\right\vert ;f_{H}\right)  \right\vert _{\left\vert n\right\vert
>1}\\
=\left\vert n\right\vert c_{2}\left[  \left\vert n\right\vert \left(
1-c_{2}\varepsilon\right)  -2\left\vert \theta\left(  f_{H}\right)
\right\vert \left(  1-\varepsilon\right)  \right]  +\left\vert n\right\vert
^{2}c_{1}c_{2}\left[  \frac{4K\left(  k_{n}\right)  }{\pi^{2}}\left[
2E\left(  k_{n}\right)  -\left(  1-k_{n}^{2}\right)  K\left(  k_{n}\right)
\right]  -1\right] \\
=\left\vert n\right\vert \frac{2\mathcal{L}_{m}}{\Phi_{0}^{2}\varepsilon
}\Delta G\left(  1;f_{H}\right)  +c_{2}\left\vert n\right\vert \left(
\left\vert n\right\vert -1\right)  \left[  \left(  1-c_{2}\varepsilon\right)
+\frac{4c_{1}K\left(  k_{n}\right)  }{\pi^{2}}\left[  2E\left(  k_{n}\right)
-\left(  1-k_{n}^{2}\right)  K\left(  k_{n}\right)  \right]  -c_{1}\right] \\
>\frac{2\mathcal{L}_{m}}{\Phi_{0}^{2}\varepsilon}\Delta G\left(
1;f_{H}\right)  ,
\end{gather*}
which was to be proved.

\section{Particular examples of soliton solutions}

\begin{figure}[ptb]
\begin{center}
$%
\begin{array}
[c]{cc}%
\includegraphics[width=0.5\textwidth]{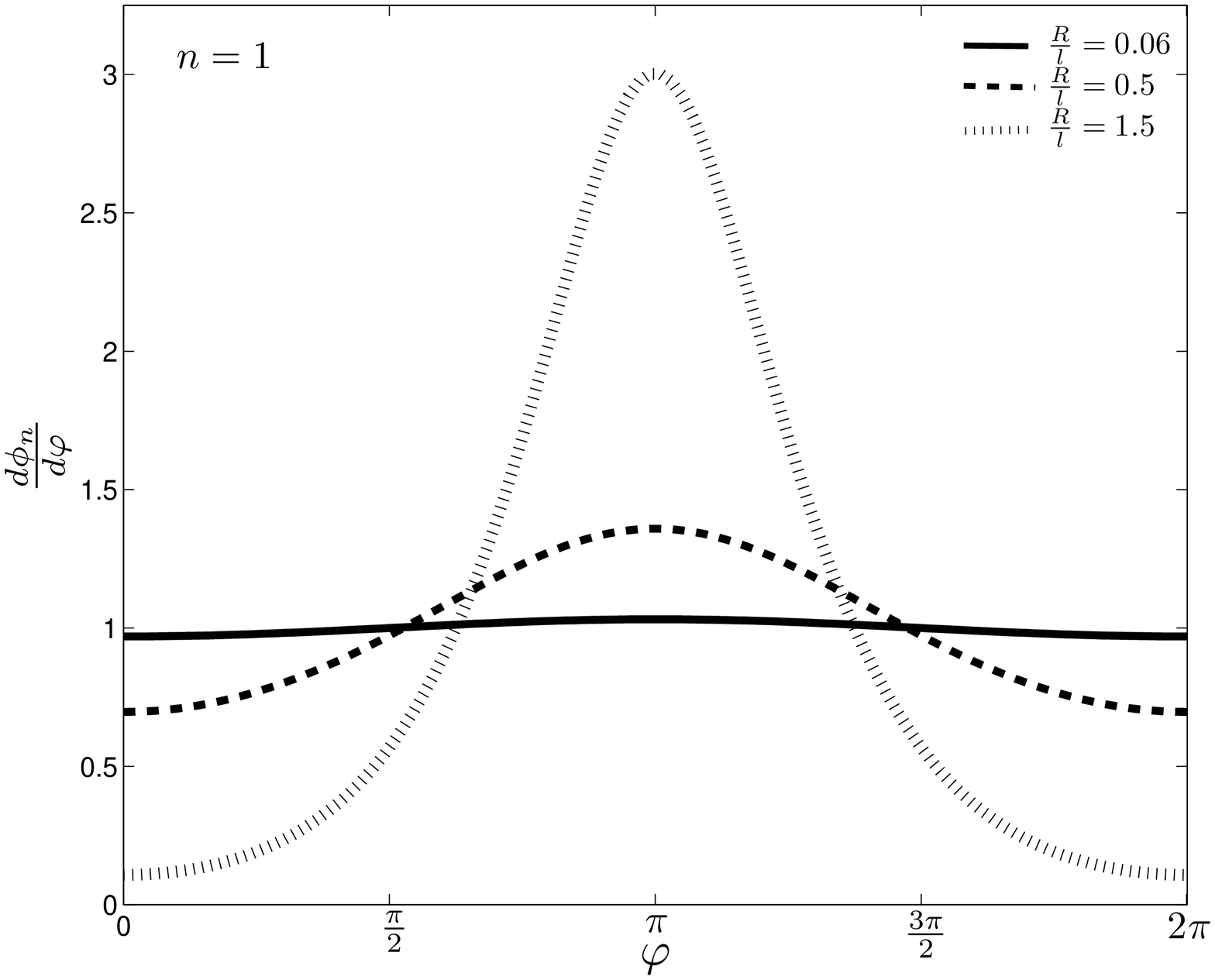} &
\includegraphics[width=0.5\textwidth]{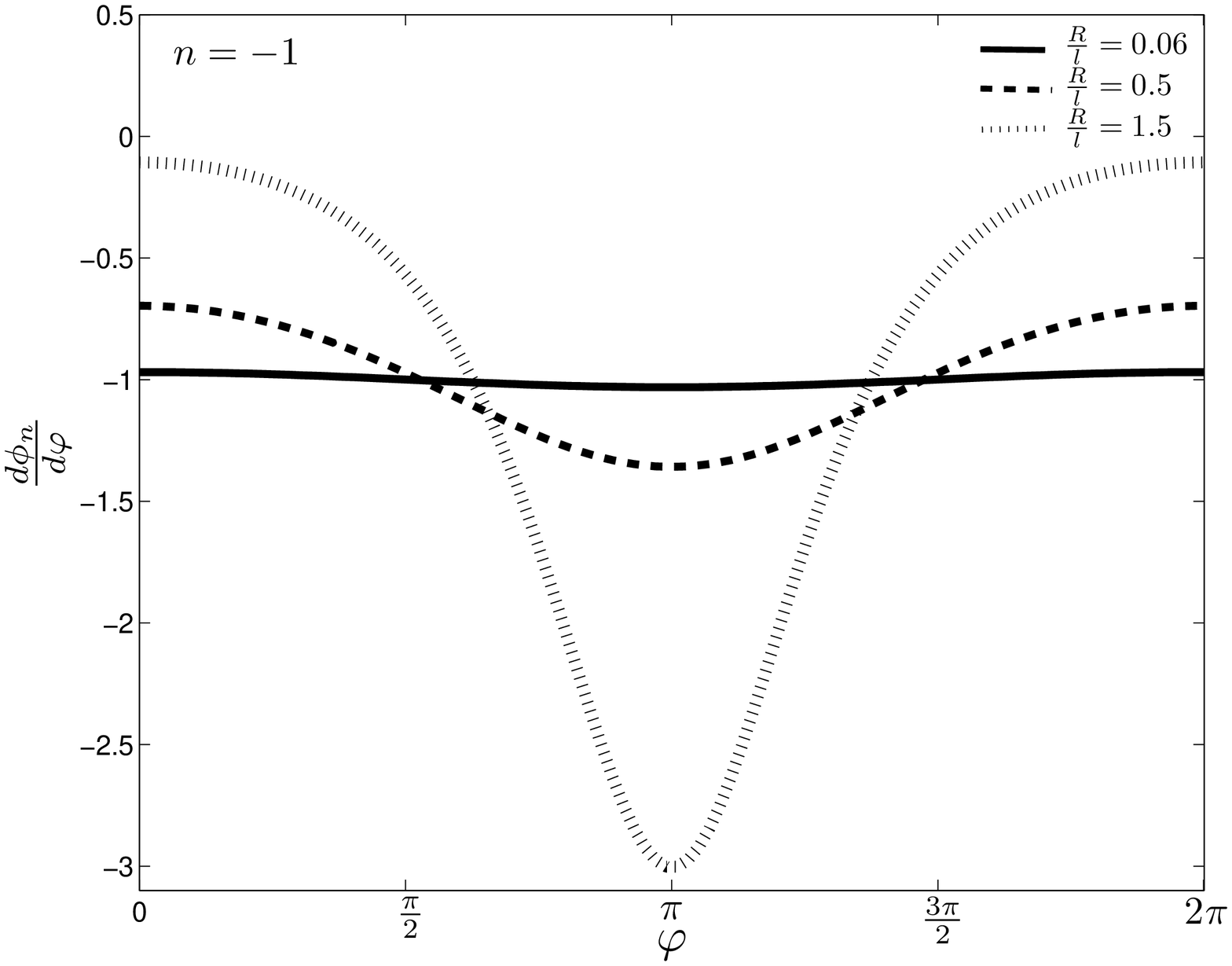}\\
\includegraphics[width=0.5\textwidth]{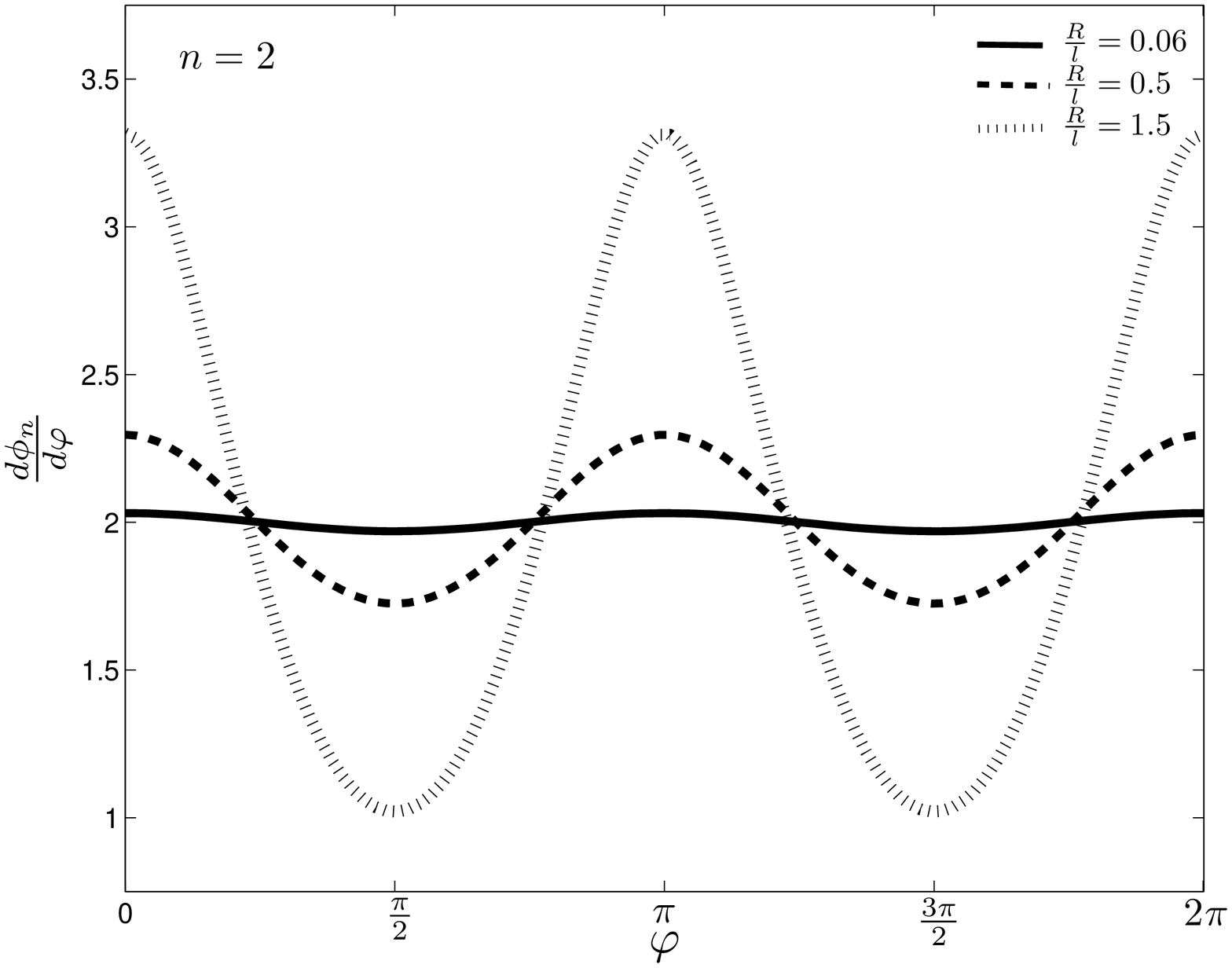} &
\includegraphics[width=0.5\textwidth]{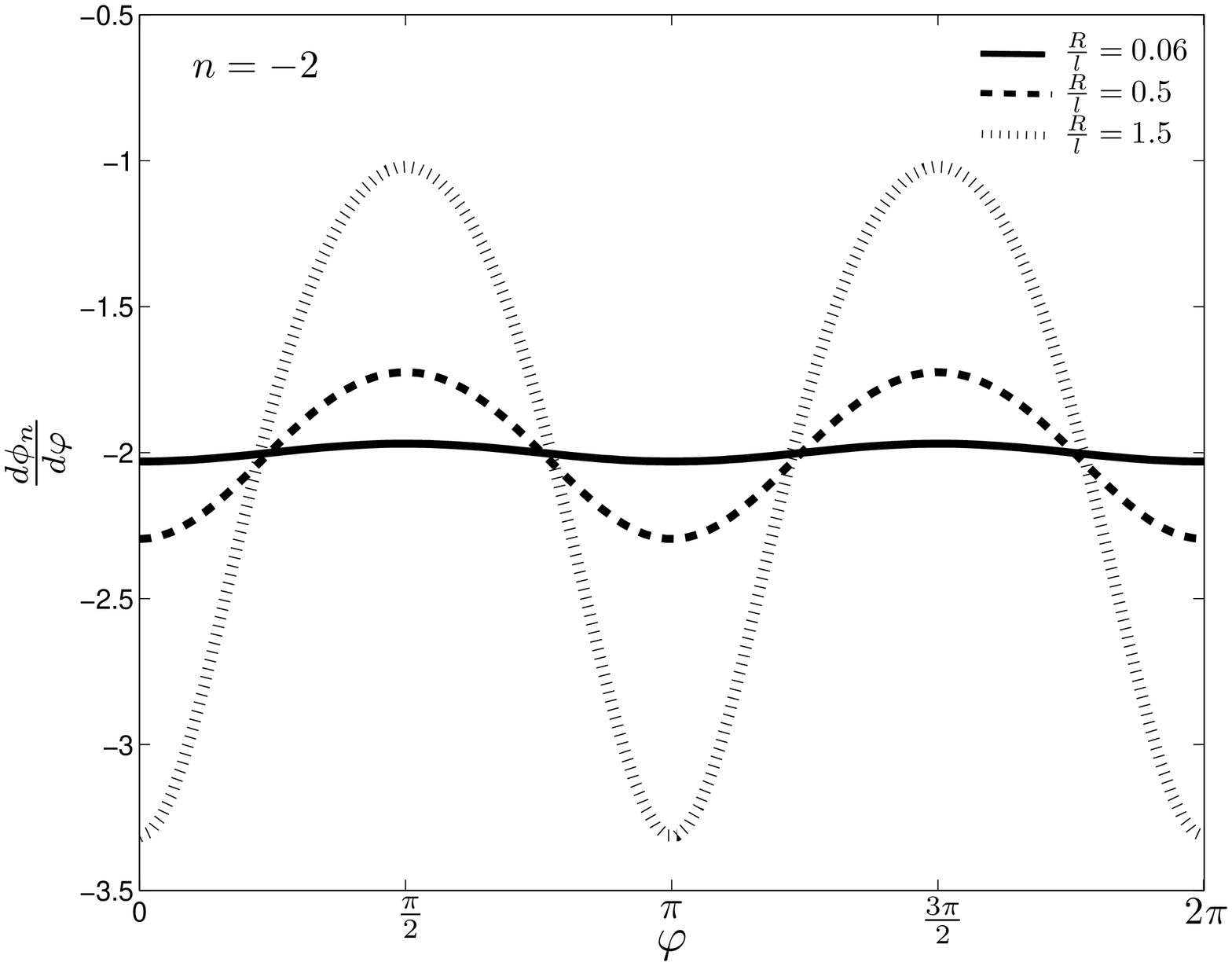}
\end{array}
$
\end{center}
\caption{Particular examples of soliton solutions ($n=\pm1,\pm2$). The
constants of integration in (\ref{3.5}) are fixed by the condition
$\varphi_{n0}=\pi$.}%
\label{fig:3}%
\end{figure}

In Fig. \ref{fig:3}, we present several different soliton solutions obtained
numerically. (For greater clarity, we plot the derivatives $\frac{d\phi_{n}%
}{d\varphi}$.) However, in two limiting cases soliton solutions possess
asymptotics in terms of elementary functions. Thus, for $\frac{R}{l}\ll1$, we
have:\cite{KG06,KG07}%
\begin{equation}
\phi_{n}\left(  \varphi\right)  \approx\frac{\left(  1+\text{sgn~}%
\gamma\right)  \pi}{2}+n\left(  \varphi-\varphi_{0}\right)  +\frac{R^{2}%
}{n^{2}l^{2}}\sin\left[  n\left(  \varphi-\varphi_{0}\right)  \right]  .
\label{c1}%
\end{equation}
The self-energy of soliton solutions (\ref{c1}) is%
\begin{equation}
F_{sol}\left(  n\right)  \approx\frac{\Phi_{0}^{2}\varepsilon}{2\mathcal{L}%
_{m}}\left\vert n\right\vert ^{2}c_{1}c_{2}\left(  1+\frac{2R^{2}}{\left\vert
n\right\vert ^{2}l^{2}}\right)  . \label{c2}%
\end{equation}
[Notice that expression (\ref{c2}) clearly illustrates the general features of
the self-energy of soliton solutions established in Section IV.]

In the opposite limiting case, when $1\ll\frac{R}{l}<\infty$, asymptotics can
be derived only for the single-soliton solutions ($\left\vert n\right\vert
=1$). Fixing the constants of integration by the condition $\varphi_{n0}=\pi$,
we get:%
\begin{equation}
\phi_{\pm1}\left(  \varphi\right)  =\frac{\left(  1+\text{sgn~}\gamma\right)
\pi}{2}\pm\left[  -\pi+4\arctan e^{\frac{R\left(  \varphi-\pi\right)  }{l}%
}+8e^{-\frac{2\pi R}{l}}\sinh\frac{R\left(  \varphi-\pi\right)  }{l}+o\left(
e^{-\frac{2\pi R}{l}}\right)  \right]  .\label{c3}%
\end{equation}
The self-energy of these solutions is%
\begin{equation}
F_{sol}\left(  \pm1\right)  =\frac{\Phi_{0}^{2}\varepsilon}{\mathcal{L}_{m}%
}\frac{4R}{\pi l}c_{1}c_{2}\left[  1+o\left(  e^{-\frac{2\pi R}{l}}\right)
\right]  .\label{c4}%
\end{equation}
Solutions (\ref{c3}) approach the well-known\cite{KG06} exact
single-soliton solutions of the static sine-Gordon equation on an infinite
interval:%
\[
\phi_{\pm1}\left(  \varphi\right)  =\frac{\left(  1+\text{sgn~}\gamma\right)
\pi}{2}\pm\left[  -\pi+4\arctan e^{\pi x}\right]  ,~x\equiv\frac{R\left(
\varphi-\pi\right)  }{\pi l}\in\left(  -\infty,+\infty\right)  .
\]


\begin{thebibliography}{99}
\bibitem {LL68}See, e.g., L. D. Landau and E. M. Lifshitz,
\textit{Electrodynamic of Continuous Media} (Pergamon Press, Oxford, 1968).
For reference purposes, we remark that the kinetic inductance is
$\mathcal{L}_{k}=\mathcal{L}_{m}\frac{1-2\varepsilon}{\varepsilon}$, as
follows from relation (\ref{2.33}). Accordingly, the expansion parameter can
be expressed as $\varepsilon=\frac{\mathcal{L}_{m}}{\mathcal{L}_{k}%
+2\mathcal{L}_{m}}\approx\frac{\mathcal{L}_{m}}{\mathcal{L}_{k}}\ll1$.
\end{thebibliography}
\end{document}